\documentclass[pra,
,tightenlines
,showkeys
,showpacs
]
{revtex4-2}

\usepackage{amsmath,amssymb}
\usepackage{bm}
\usepackage[dvipdfmx]{graphicx}
\usepackage{ascmac} 
\usepackage{fancybox}
\usepackage{float}
\usepackage{enumerate}
\usepackage{color}
\usepackage{amsthm}
\usepackage{mathrsfs}
\usepackage{comment}
\usepackage{ulem}
\allowdisplaybreaks[1]
\usepackage{comment}
\newtheorem*{th.}{Theorem}

\newcommand{\vsigma}{\vec{\sigma}}

\newcommand{\tr}{{\rm Tr}}

\newcommand{\ket}[1]{|#1\rangle}

\newcommand{\am}{{\rm am}}

\begin{document}
	\title{Optimal quantum controls robust against detuning error}
	\author{Shingo Kukita$^{1)}$}
	\email{kukita@nda.ac.jp}
	\author{Haruki Kiya$^{2)}$}
	\email{kiya.haruki@kindai.ac.jp }
	\author{Yasushi Kondo$^{2)}$}
	\email{ykondo@kindai.ac.jp}
	\affiliation{$^{1)}$Department of Computer Science, National Defence Academy of Japan, 1-10-20, Hashirimizu, Yokosuka, Japan}
	\affiliation{$^{2)}$Department of Physics, Kindai University, Higashi-Osaka 577-8502, Japan}
	
\begin{abstract}

Precise control of quantum systems is one of the most important milestones for achieving practical quantum technologies, such as computation, sensing, and communication.
Several factors deteriorate the control precision and thus their suppression is strongly demanded.
One of the dominant factors is systematic errors, which are caused by discord between an expected parameter in control and its actual value.
Error-robust control sequences, known as composite pulses, have been invented in the field of nuclear magnetic resonance (NMR).
These sequences mainly focus on the suppression of errors in one-qubit control.
The one-qubit control, which is the most fundamental in a wide range of quantum technologies, often suffers from detuning error.
As there are many possible control sequences robust against the detuning error, it will practically be important to find ``optimal" robust controls with respect to several cost functions such as time required for operation, and pulse-area during the operation, which corresponds to the energy necessary for control.
In this paper, we utilize the Pontryagin's maximum principle (PMP), a tool for solving optimization problems under inequality constraints, to solve the time and pulse-area optimization problems.
We analytically obtain pulse-area optimal controls robust against the detuning error.
Moreover, we found that short-CORPSE, which is the shortest known composite pulse so far, is a probable candidate of the time optimal solution according to the PMP.
We evaluate the performance of the pulse-area optimal robust control and the short-CORPSE, comparing with that of the direct operation.

\end{abstract}

\maketitle
\section{introduction}
Achieving precise control of quantum systems is one of the most important steps towards realization of future quantum technologies, such as quantum communication \cite{ekert1991quantum,gisin2007quantum,chen2021integrated}, sensing \cite{helstrom1976quantum,caves1981quantum,holevo2011probabilistic}, computing \cite{Nielsen2000,bennett2000quantum,nakahara2008quantum}.
Such a precise control is particularly desired for noisy intermediate scale quantum (NISQ) computers \cite{preskill2018quantum} because the NISQ computers cannot implement quantum error correction:
imprecision in each control is directly reflected in the performance of computations.
Numerous attempts have been performed to ameliorate the precision of quantum control \cite{RevModPhys.76.1037,PhysRevLett.99.036403,bason2012high,chang2014band,spiteri2018quantum,levy2018noise}.
One-qubit control, which is a foundation of many practical quantum technologies, has intensively been investigated to improve its performance \cite{PhysRevA.73.022332,PhysRevLett.111.050404,PhysRevA.94.032323,yang2019achieving}.

Systematic errors in devices are a major factor deteriorating the performance of quantum control.
Slow temporal fluctuations of parameters in a system to be considered, which cause systematic errors, are inevitable in real devices.
Although effects of decoherence are considered to be more dominant than such errors in current quantum devices, several studies have reported that systematic errors also have an impact on the realization of quantum technologies \cite{PhysRevLett.87.227901,PhysRevA.94.042338,PhysRevA.99.022325,PhysRevLett.129.250503}.
Software suppression of their effects will be necessary because it is difficult to mitigate them only by hardware calibration.
Error-robust control sequences compensating for the effects of systematic errors, which are known as ``composite pulses", have been developed in the field of nuclear magnetic resonance (NMR) \cite{counsell1985analytical,levitt1986composite,claridge2016high}.
These error-robust sequences are composed of several operations such that errors in the operations are canceled each other.
Its efficacy has been demonstrated in several experimental testbeds; not only NMR, but also nitrogen-vacancy center \cite{said2009robust}, superconducting qubits \cite{collin2004nmr,PhysRevApplied.18.034062}, and ion traps \cite{timoney2008error}.

One-qubit control typically suffers from two systematic errors: amplitude error (or pulse length error in NMR) and detuning error (off-resonance error).
Many error-robust controls have been constructed to attain robustness against each of these errors; e.g., SK1 \cite{brown2004arbitrarily}, BB1 \cite{wimperis1994broadband}, and SCROFULOUS \cite{cummins2003tackling} against the amplitude error, and the CORPSE family \cite{cummins2000use} against the detuning error.
Furthermore, several studies consider control sequences simultaneously robust against both errors \cite{ryan2010robust,bando2020concatenated,jones2013designing,doi:10.7566/JPSJ.91.104001}.

There are many possible robust controls even once the error to be suppressed is fixed. 
Therefore, one should discuss ``optimal" robust controls with respect to several costs, such as time required for the whole operations or pulse area.
Robust controls typically have long operation time and large pulse area because they suppress the effects of error by utilizing the redundancy of quantum dynamics achieving a target operation.
Several studies have addressed optimal controls robust against certain errors \cite{KOBZAR2004236,6189046,PhysRevLett.111.050404,PhysRevA.95.063403,PhysRevA.100.023415,PhysRevLett.125.250403,Ansel_2021,PhysRevA.106.052608}.
They mainly focus on robust controls implementing certain fixed gates such as the NOT gate, or focus on state transfer, i.e., robust control with fixed initial states.

In this paper, we consider time and pulse-area optimal controls robust against the detuning error utilizing the Pontryagin's maximum principle (PMP), a mathematical theory for solving optimization problems under inequality constraints \cite{pontryagin2018mathematical}.
We analytically found the pulse-area optimal robust control for a wide range of target operations under some requirements motivated by experimental implementation.
Also, this control has the gate-level robustness; i.e., independent of the choice of initial states.
Furthermore, we show that the short-CORPSE in the CORPSE family, which was the shortest known optimal robust sequence so far, is a probable candidate of the time optimal robust control although we could not provide a complete proof that it is the time optimal.
Both pulse-area optimal sequence and short-CORPSE have similar geometric structure of the dynamics:
in the Bloch sphere representation, they are sequences of coaxial rotations and have ``switchback" behaviour, which is often used to compensate for the detuning error \cite{cummins2000use,doi:10.7566/JPSJ.91.104001}.
We compare their performance with the direct target operation, of which dynamics is implemented by a constant Hamiltonian.

This paper is organized as follows.
In Sec.~\ref{sec:2}, we provide a brief review on the construction of robust controls following the definition of the composite pulses.
Section~\ref{sec:III} provides how to solve the optimization problems with constraints of the detuning robustness and obtain the pulse-area optimal control and the probable time optimal one.
We then evaluate their overheads comparing with the corresponding direct operation in Sec.~\ref{sec:IIII}.
Section~\ref{sec:IIIII} is devoted to the summary of this work.
We set $\hbar=1$ throughout this paper.
 
\section{brief review of error-robust control}
\label{sec:2}

In this section we explain the essence of error-robust control following the concept of the composite pulses.
We first review basic facts in qubit dynamics, and then consider dynamics with unavoidable errors.
Hereinafter, ${\mathbb I}_{2}$ denotes the $2\times 2$ identity matrix and $\sigma_{x,y,z}$ are Pauli matrices:
\begin{equation}
{\mathbb I}_{2}=
\begin{pmatrix}
1&0\\
0&1
\end{pmatrix}
,~~\sigma_{x}=
\begin{pmatrix}
0&1\\
1&0
\end{pmatrix}
,~~\sigma_{y}=
\begin{pmatrix}
0&-i\\
i&0
\end{pmatrix}
,~~\sigma_{z}=
\begin{pmatrix}
1&0\\
0&-1
\end{pmatrix}
.
\end{equation}
Also, we shall take $t=0$ and $T$ as the initial and termination time of dynamics, respectively.

\subsection{qubit dynamics} 

A qubit has a state in the two-dimensional Hilbert space spanned by $\{\ket{0},\ket{1}\}$.
The dynamics of the state is governed by the following Schr\"odinger equation:
\begin{equation}
\frac{d}{d t} \ket{\psi(t)}=-i H(t)\ket{\psi(t)},
\label{eq:schroedinger}
\end{equation}
where $H(t)$ is a $2\times 2$ Hermitian matrix called Hamiltonian.
Introducing a unitary matrix $U(t)\in SU(2)$ such that $\ket{\psi(t)}=U(t)\ket{\psi_{0}}$ where $\ket{\psi_{0}}$ is an initial state,
we can derive the equation for the unitary matrix:
\begin{equation}
\frac{d}{d t} U(t)=-i H(t)U(t).
\end{equation}
The corresponding initial condition is $U(0)={\mathbb I}_{2}$.
The solution of the above equation is formally given by
\begin{equation}
U(t)={\cal T}\exp\left(-i \int^{t}_{0} d t'  H(t')\right),
\end{equation}
where ${\cal T}$ represents the time-ordering product.

Any unitary matrix can be written in the ``rotational representation" $R(\theta,\vec{k})$ defined as
\begin{equation}
R(\theta,\vec{k}):=\cos(\theta/2){\mathbb I}_{2}-i \sin(\theta/2)\vec{k}\cdot\vsigma,
\label{eq:rot_rep}
\end{equation}
where $\vec{k}$ is a three-dimensional unit vector, and $\vsigma=(\sigma_{x},\sigma_{y},\sigma_{z})$.
If we represent a state of the qubit by a point on the two-dimensional unit sphere (Bloch sphere), $R(\theta,\vec{k})$ corresponds to the rotation with the axis $\vec{k}$ and the angle $\theta$; therefore, we call it the rotational representation.

An essential task in qubit control is to achieve a target unitary at $t=T$, i.e., $U(T)=R(\theta_{f},\vec{k}_{f})$ while adjusting $H(t)$ under some requirements.
Obviously, if we have no requirement on the control, the simplest Hamiltonian to achieve the target operation $R(\theta_{f},\vec{k}_{f})$ is
\begin{equation}
H(t)=\frac{\theta_{f}}{T}\vec{k}_{f}\cdot \frac{\vsigma}{2},
\label{eq:trivial}
\end{equation}
which is constant during $0\leq t \leq T$.
In practical situations, however, there may be some requirements, e.g.,
\begin{itemize}
\item we cannot generate the $z$ component of $H(t)$: $\tr\left(H(t)\sigma_{z}\right)=0$,
\item we must guarantee error robustness, which is a main scope of this paper, and
\item the ``strength" of the (control) Hamiltonian will also be restricted : $\sqrt{\tr\left(H(t)H(t)\right)}\leq \Omega$.
\end{itemize}
Furthermore, under the above constraints, we often need to optimize some quantities, such as the termination time $T$ required to achieve the target unitary or the total energy consumption during the control.

\subsection{composite quantum gate}

The dynamics of the qubit sometimes suffers from systematic errors.
Provided that we apply a control Hamiltonian $H_{\rm cont}(t)$ to the qubit, its actual dynamics is governed by the faulty Hamiltonian:
\begin{equation}
H_{\rm total}(t)=H_{\rm cont}(t)+f H_{\rm error}(t),
\end{equation}
where $f$ is a small constant representing the (signed) strength of the error Hamiltonian.
We assume that the form of $H_{\rm error}(t)$ is known while $f$ is unknown.
The corresponding dynamics of the unitary is given by
\begin{equation}
\frac{d}{d t}U(t)=-i\left(H_{\rm cont}(t)+f H_{\rm error}(t)\right)U(t).
\end{equation}
Expanding the unitary $U(t)$ up to the first order of $f$, i.e., $U(t)=U^{(0)}(t)+f U^{(1)}(t)+{\cal O}(f^{2})$,
we obtain the order-by-order equations:
\begin{equation}
\frac{d}{d t}U^{(0)}(t)=-iH_{\rm cont}(t)U^{(0)}(t),~~\frac{d}{d t}U^{(1)}(t)=-i\left(H_{\rm cont}(t)U^{(1)}(t)+H_{\rm error}(t)U^{(0)}(t)\right).
\label{eq:orderbyorder}
\end{equation}
Note that $U^{(1)}(t)$ is neither unitary nor Hermitian while $U^{(0)}(t)$ is unitary.
Now, let us try to design an error-robust control reaching the target operation $R(\theta_{f},\vec{k}_{f})$ after the dynamics.
This is formulated by the boundary conditions for $U^{(0)}(t)$ and $U^{(1)}(t)$ as
\begin{equation}
U^{(0)}(0)={\mathbb I}_{2},~~U^{(1)}(0)=0,~~U^{(0)}(T)=R(\theta_{f},\vec{k}_{f}),~~U^{(1)}(T)=0.
\label{eq:robust}
\end{equation} 
Thus, we can achieve the target operation in an error-robust way:
\begin{equation}
U(T)=U^{(0)}(T)+f U^{(1)}(T)+{\cal O}(f^{2})=R(\theta_{f},\vec{k}_{f})+{\cal O}(f^{2}).
\end{equation}
We call a scheduling of the control Hamiltonian $H_{\rm cont}(t)$ achieving the boundary conditions~(\ref{eq:robust}), or the corresponding $0$-th unitary $U^{(0)}(t)$, an (first-order) error-robust control against $H_{\rm error}(t)$, which is essentially the same as ``composite pulse" in NMR.
There are in general many possibilities of the control Hamiltonian achieving the conditions.

Note that the above construction is not the unique way to attain the error-robustness; e.g., if we know the error strength $f$, there should be more efficient ways for the error suppression.
We can also utilize ancillary state spaces if a qubit is spanned by the lowest and first-excited states in a multi-level system.
In this paper, however, we only consider the above type of error-robust controls.

A typical error in one-qubit control is {\it detuning error} defined as
\begin{equation}
H_{\rm error}(t)=\frac{\sigma_{z}}{2},
\label{eq:direct}
\end{equation}
which arises from discord between the Larmor frequency and the frequency of an applied field.
Let us consider the control Hamiltonian achieving a target operation in the form of
\begin{equation}
\bar{R}(\theta_{f},\phi_{f})=R\left(\theta_{f},\vec{n}_{\phi_{f}}=(\cos\phi_{f},\sin\phi_{f},0)\right),
\end{equation}
in an detuning-robust way.
The CORPSE family is a three-parameter family of detuning-robust controls achieving $\bar{R}(\theta_{f},\phi_{f})$ for arbitrary $\theta_{f}$ and $\phi_{f}$ \cite{cummins2000use}.
The short-CORPSE has the shortest operation time in this family, and its sequence is described by the following piecewise Hamiltonian:
\begin{equation}
H_{\rm cont}(t)=
\begin{cases}
-\Omega \vec{n}_{\phi_{f}}\cdot\frac{\vsigma}{2},&0\leq \Omega t \leq \theta^{\rm (SC)}_{1},\\
\Omega \vec{n}_{\phi_{f}}\cdot\frac{\vsigma}{2},&\theta^{\rm (SC)}_{1} \leq \Omega t \leq \theta^{\rm (SC)}_{1}+\theta^{\rm (SC)}_{2},\\
-\Omega \vec{n}_{\phi_{f}}\cdot\frac{\vsigma}{2},&\theta^{\rm (SC)}_{1}+\theta^{\rm (SC)}_{2}\leq \Omega t \leq 2 \theta^{\rm (SC)}_{1}+\theta^{\rm (SC)}_{2},
\end{cases}
\end{equation}
where $\theta^{\rm (SC)}_{1}=\pi-\kappa-\theta_{f}/2$, $\theta^{\rm (SC)}_{2}=2\pi-2\kappa$, and $\kappa=\arcsin\left(\sin(\theta_{f}/2)/2\right)$.
As this sequence is a coaxial rotation, the $0$-th unitary matrix $U^{(0)}(t)$ during the dynamics is simply described as
\begin{equation}
U^{(0)}(t)=\exp\left(-i \theta(t) \vec{n}_{\phi_{f}}\cdot\frac{\vsigma}{2} \right),
\end{equation}
where $\theta(t)$ satisfies
\begin{equation}
\theta(t)=
\begin{cases}
- \Omega t,&0\leq \Omega t \leq \theta^{\rm (SC)}_{1},\\
 \Omega t-2\theta^{\rm (SC)}_{1},&\theta^{\rm (SC)}_{1} \leq \Omega t \leq \theta^{\rm (SC)}_{1}+\theta^{\rm (SC)}_{2},\\
- \Omega t+\theta^{\rm (SC)}_{2},&\theta^{\rm (SC)}_{1}+\theta^{\rm (SC)}_{2}\leq \Omega t \leq 2 \theta^{\rm (SC)}_{1}+\theta^{\rm (SC)}_{2}.
\end{cases}
\label{eq:shortCORPSE}
\end{equation}
We should note that $\bar{R}(2\pi -\theta_{f},\phi_{f})$ is utilized instead of $\bar{R}(\theta_{f},\phi_{f})$ for $\theta_{f}\leq\pi$ 
in its original version because they are equivalent up to a global phase and the former has a shorter operation time.

\subsection{time and pulse-area optimization}

Error robust controls tend to have a longer operation time than the direct operation (\ref{eq:direct}).
A natural question is, how short operation time $T$ can be realized while error robustness being satisfied.
This is important from the viewpoint of noise immunity; if a control has a long operation time, noise from an environment causes decoherence of the qubit, and thus hinders precise control.
Thus, we consider the optimization problem of $T$.
For later convenience, we rewrite the termination time $T$ in the integral form:
\begin{equation}
T=\int^{T}_{0} d t = \int^{T}_{0} d t f^{(t)}\left(H_{\rm cont}(t)\right),
\end{equation}
although $f^{(t)}\left(H_{\rm cont}(t)\right)$ is actually $1$.
Hereinafter, we refer to $T$ as $L^{(t)}$ when we mean by it the cost function to be optimized while we keep using $T$ for the upper limit of integration and the termination time of the dynamics.

Note that ``physical" time has no crucial meaning.
If we could unlimitedly increase the strength of the control field, the operation would be done within an infinitesimal time, and the control is not affected by the detuning error.
Hence, we somehow need to introduce a reference value of time.
In this paper, we shall consider optimization problems while fixing a maximum strength $\Omega$ of the control field, which is reasonable in experiments, and treat its inverse as the time reference.
The time variable $t$ is then non-dimensionalized by $\Omega$: $t\leftarrow\Omega t$.
Accordingly, the strength of the Hamiltonian are renormalized as
\begin{equation}
\sqrt{\tr\left(H_{\rm cont}(t)H_{\rm cont}(t)\right)}\leftarrow\sqrt{\tr\left(H_{\rm cont}(t)H_{\rm cont}(t)\right)}/\Omega,~~f\leftarrow f/\Omega.
\end{equation}
Thus, the strength of the control Hamiltonian is bounded by $1$ rather than $\Omega$.
 
Another important ``cost" in qubit control is the pulse area $L^{(p)}$, which is defined as
\begin{equation}
L^{(p)}=\int^{T}_{0} d t f^{(p)}\left(H_{\rm cont}(t)\right)=\int^{T}_{0} d t \sqrt{\tr\left(H_{\rm cont}(t)H_{\rm cont}(t)\right)}.
\end{equation}
This quantity represents the accumulation of the strength of the control Hamiltonian during the dynamics, and it is preferable that the control takes a smaller $L^{(p)}$.
We note that when we consider an error that is invariant under time rescaling, optimization of the control with respect to $L^{(p)}$ is equivalent to that with respect to $L^{(t)}$.
The amplitude error, another typical error in one-qubit control has this property \cite{PhysRevLett.125.250403}.
Meanwhile, the detuning error does not, and thus we should independently discuss the time and pulse-area optimizations for detuning-robust controls.

It is rather convenient to optimize
\begin{equation}
L^{(e)}=\int^{T}_{0} d t f^{(e)}\left(H_{\rm cont}(t)\right)=\int^{T}_{0} d t \tr\left(H_{\rm cont}(t)H_{\rm cont}(t)\right),
\end{equation}
than directly dealing with the pulse-area optimization.
These optimization problems have the same solution because $L^{(e)}$ monotonically increases as $L^{(p)}$ does.
The $L^{(e)}$-optimization is, however, easier to treat: its integrand does not contain a square root.

\section{formulation in pontryagin's maximum principle}
\label{sec:III}

We first assume that the target operation is $\bar{R}(\theta_{f},0)$, which is the rotation along the $x$ axis while $\theta_{f}\in[0,2\pi]$ is arbitrary.
Extension to target operations in the form of $\bar{R}(\theta_{f},\phi_{f})$ is almost trivial because the system to be considered is invariant under the rotation along the $z$ axis.
Moreover, the control Hamiltonian is assumed to have no $z$ component throughout the dynamics:
\begin{equation}
H_{\rm cont}(t)=  \omega_{x}(t)\frac{\sigma_{x}}{2}+\omega_{y}(t)\frac{\sigma_{y}}{2}.
\end{equation}
If we could freely control the $z$ component, the detuning error, which arises from the incompletion of control of the $z$ component, would never be generated. 
As aforementioned, the strength of the control Hamiltonian at each time is assumed to be bounded by $1$, i.e.,
\begin{equation}
\tr \left(H_{\rm cont}(t)H_{\rm cont}(t)\right)=\left(\omega_{x}(t)\right)^{2}+\left(\omega_{y}(t)\right)^{2}\leq 1.
\end{equation}
Under these premises we try to find time and pulse-area optimal controls robust against the detuning error.

We reformulate the above problem in the PMP, of which details are given in Appendix~\ref{sec:pmp}, because one cannot apply the ordinary variational principle to optimization problems with inequality constraints as above.
We begin with introducing an auxiliary scalar variable:
\begin{equation}
\upsilon^{(\alpha)}(t)=\int^{t}_{0} d t' f^{(\alpha)}\left(H_{\rm cont}(t')\right)=\int^{t}_{0} d t' f^{(\alpha)}\left(\omega_{x}(t'),\omega_{y}(t')\right),
\end{equation}
where $\alpha$ represents $t$ (time optimization case) or $e$ (energy optimization case).
This variable satisfies the dynamical equation,
\begin{equation}
\frac{d \upsilon^{(\alpha)}(t)}{d t}=f^{(\alpha)}\left(\omega_{x}(t),\omega_{y}(t)\right),
\label{eq:aux}
\end{equation}
with the boundary conditions, $\upsilon^{(\alpha)}(0)=0$, and $\upsilon^{(\alpha)}(T)=L^{(\alpha)}$.
Furthermore, we introduce the ``conjugate momenta", $P_{(0)}(t),P_{(1)}(t)$, which are matrices, and $\pi_{(\alpha)}(t)\in {\mathbb R}$, through the differential equations,
\begin{equation}
\frac{d}{d t}P_{(0)}(t)=i\left(P_{0}(t)H_{\rm cont}(t)+P_{1}(t)H_{\rm error}\right),~~\frac{d}{d t}P_{(1)}(t)=i P_{(1)}(t)H_{\rm cont}(t),~~\frac{d \pi_{(\alpha)}(t)}{d t}=0,
\label{eq:momenta}
\end{equation}
where we use the notation $H_{\rm error}:=\sigma_{z}/2$ for simplicity.
Although the matrices $P^{(0)}$, $P^{(1)}$, and $U^{(1)}$ are not necessarily unitary,
one can expand them in the following forms:
\begin{align}
U^{(0,1)}=&u^{(0,1)}_{I}{\mathbb I}_{2}-i\left(u^{(0,1)}_{x}\sigma_{x}+u^{(0,1)}_{y}\sigma_{y}+u^{(0,1)}_{z}\sigma_{z}\right),~~u_{I,x,y,z}^{(0,1)}\in {\mathbb R},\nonumber\\
P_{(0,1)}=&p_{(0,1)}^{I}{\mathbb I}_{2}+i\left(p_{(0,1)}^{x}\sigma_{x}+p_{(0,1)}^{y}\sigma_{y}+p_{(0,1)}^{z}\sigma_{z}\right),~~p^{I,x,y,z}_{(0,1)}\in {\mathbb R},
\label{eq:basisexp}
\end{align}
where all the coefficients are real because of the dynamical equations they obey.

We define
\begin{align}
K^{(\alpha)}(U^{(0,1)},P_{(0,1)},\pi_{(\alpha)};\omega_{x},\omega_{y})=\pi_{(\alpha)} f^{(\alpha)}\left(\omega_{x},\omega_{y}\right)-i&\tr\left(P_{(0)}H_{\rm cont}U^{(0)}\right)\nonumber\\
&-i\tr\left(P_{(1)}\left(H_{\rm cont}U^{(1)}+H_{\rm error}U^{(0)}\right)\right),
\label{eq:ponthamiltonian}
\end{align}
which is easily proved to be always real despite the imaginary unit.
Using this quantity, we rewrite Eqs.~(\ref{eq:orderbyorder}), (\ref{eq:aux}), and (\ref{eq:momenta}) as
\begin{equation}
\frac{d u^{(0,1)}_{I,x,y,z}}{d t}=\frac{\partial K^{(\alpha)}}{\partial p^{I,x,y,z}_{(0,1)}},~~\frac{d \upsilon^{(\alpha)}}{d t}=\frac{\partial K^{(\alpha)}}{\partial \pi_{(\alpha)}},~~
\frac{d p_{(0,1)}^{I,x,y,z}}{d t}=-\frac{\partial K^{(\alpha)}}{\partial u_{I,x,y,z}^{(0,1)}},~~\frac{d \pi_{(\alpha)}}{d t}=-\frac{\partial K^{(\alpha)}}{\partial \upsilon^{(\alpha)}},
\end{equation}
just like classical Hamilton's equations of motion, and hence we call $K^{(\alpha)}$ ``Hamiltonian".
The matrix parts of the equations are formally summarized as
\begin{equation}
\frac{d U^{(0)}}{d t}=\frac{\partial K^{(\alpha)}}{\partial P_{(0)}},~~\frac{d U^{(1)}}{d t}=\frac{\partial K^{(\alpha)}}{\partial P_{(1)}},~~
\frac{d P_{(0)}}{d t}=-\frac{\partial  K^{(\alpha)}}{\partial U^{(0)}},~~\frac{d P_{(1)}}{d t}=-\frac{\partial  K^{(\alpha)}}{\partial U^{(1)}}.
\end{equation}

The PMP states that (if exist) a control $\omega^{*}_{x,y}(t)$ optimizing $L^{(\alpha)}$ satisfies
\begin{equation}
(\omega^{*}_{x},\omega^{*}_{y})=\underset{\omega^{2}_{x}+\omega^{2}_{y}\leq 1}{\rm argmax} K^{(\alpha)}(U^{(0,1)},P_{(0,1)},\pi_{(\alpha)};\omega_{x},\omega_{y}),
\label{eq:argmax}
\end{equation}
at each time in $0\leq t \leq T$.
Here the ${\rm argmax}$ function is the set of the inputs at which a function is maximized.
Solving Eqs.~(\ref{eq:orderbyorder}), (\ref{eq:aux}), and (\ref{eq:momenta}) with this control and the boundary conditions, one obtain a candidate of the optimal solution $U^{*(0)}(t)$, $U^{*(1)}(t)$, and the optimized cost function $L^{(\alpha)}$.
We should emphasize that this statement is a necessary condition for the optimal solution.
When we obtain several candidates satisfying the above condition, we should compare their cost function and decide which is optimal.

To simplify the problem, we consider other variables (matrices) instead of $P_{(0)}$ and $P_{(1)}$:
\begin{equation}
\Gamma=U^{(0)}P_{0}+U^{(1)}P_{(1)},~~\Delta=U^{(0)}P_{(1)},
\end{equation}
which are analogues of angular momenta.
Thus, the Hamiltonian is simplified as
\begin{equation}
K^{(\alpha)}(\Gamma,\Delta,\pi_{(\alpha)};\omega_{x},\omega_{y})=\pi_{(\alpha)} f^{(\alpha)}\left(\omega_{x},\omega_{y}\right)-i \tr\left(H_{\rm cont}\Gamma\right)-i\tr\left(H_{\rm error}\Delta\right).
\end{equation}
Also, one can find that these matrices obey the equations,
\begin{equation}
\frac{d \Gamma}{d t}=-i [H_{\rm cont},\Gamma]-i[H_{\rm error},\Delta],~~\frac{d \Delta}{d t}=-i [H_{\rm cont},\Delta],
\end{equation}
where $[\bullet,\bullet]$ is the commutator.
Solving these equation with the dynamics of $U^{(0,1)}$,
we obtain candidates of the optimal solution.
Similarly to the other variables, $\Gamma$ and $\Delta$ can be expanded in the Pauli-basis form,
\begin{align}
\Gamma=&\gamma_{I}{\mathbb I}_{2}-i\left(\gamma_{x}\sigma_{x}+\gamma_{y}\sigma_{y}+\gamma_{z}\sigma_{z}\right),~~\gamma_{I,x,y,z}\in {\mathbb R},\nonumber\\
\Delta=&\delta_{I}{\mathbb I}_{2}-i\left(\delta_{x}\sigma_{x}+\delta_{y}\sigma_{y}+\delta_{z}\sigma_{z}\right),~~\delta_{I,x,y,z}\in {\mathbb R},
\end{align}
whereby the Hamiltonian is rewritten as
\begin{equation}
K^{(\alpha)}(\gamma_{x,y},\delta_{z},\pi_{(\alpha)};\omega_{x},\omega_{y})=\pi_{(\alpha)} f^{(\alpha)}\left(\omega_{x},\omega_{y}\right)-\left(\omega_{x}\gamma_{x}+\omega_{y}\gamma_{y}\right)-\delta_{z}.
\end{equation}

Let us find $(\omega^{*}_{x},\omega^{*}_{y})$ maximizing $K^{(\alpha)}$.
In the $L^{(t)}$-optimization with $f^{(t)}(\omega_{x},\omega_{y})=1$, we obtain
\begin{equation}
(\omega^{*}_{x},\omega^{*}_{y})=\underset{\omega^{2}_{x}+\omega^{2}_{y}\leq 1}{\rm argmax}\left(\pi_{(t)}-\omega_{x}\gamma_{x}-\omega_{y}\gamma_{y}-\delta_{z}\right)=- \left(\bar{\gamma}_{x},\bar{\gamma}_{y}\right),
\label{eq:time}
\end{equation}
where $\bar{\gamma}_{x,y}:=\gamma_{x,y}/\sqrt{\gamma^{2}_{x}+\gamma^{2}_{y}}$.
In the $L^{(e)}$-optimization, we take $\pi_{(e)}=-1/2$, and obtain
\begin{equation}
(\omega^{*}_{x},\omega^{*}_{y})=\underset{\omega^{2}_{x}+\omega^{2}_{y}\leq 1}{\rm argmax}\left(-\frac{\omega^{2}_{x}+\omega^{2}_{y}}{2}-\omega_{x}\gamma_{x}-\omega_{y}\gamma_{y}-\delta_{z}\right)=-\left(\gamma_{x},\gamma_{y}\right),
\label{eq:optimal_pa}
\end{equation}
Here we assume $\gamma_{x}^{2}+\gamma_{y}^{2}\leq 1$, which is justified after calculations below.
In what follows, we explicitly solve the dynamics of $U^{(0,1)}$ with this optimal control field for the time and pulse-area optimal cases.

\subsection{pulse-area optimization}
\label{sec:III:pulse}

We should solve the pulse-area optimization first;
as shown below, the equations are rather simple in this case and the optimal dynamics is uniquely determined.
The (possibly) optimal control Hamiltonian is given by
\begin{equation}
H_{\rm cont}(t)=-\left(\gamma_{x}\frac{\sigma_{x}}{2}+\gamma_{y}\frac{\sigma_{y}}{2}\right).
\end{equation}
See also Eq.~(\ref{eq:optimal_pa}).
Writing down all the equations for the variables $U^{(0,1)}$, $\Gamma$, and $\Delta$,
we obtain
\begin{align}
\dot{u}^{(0)}_{I}=&\frac{\gamma_{x}}{2}u^{(0)}_{x}+\frac{\gamma_{y}}{2}u^{(0)}_{y},~~\dot{u}^{(0)}_{x}=-\frac{\gamma_{y}}{2}u^{(0)}_{z}-\frac{\gamma_{x}}{2}u^{(0)}_{I},~~
\dot{u}^{(0)}_{y}=\frac{\gamma_{x}}{2}u^{(0)}_{z}-\frac{\gamma_{y}}{2}u^{(0)}_{I},~~\dot{u}^{(0)}_{z}=-\frac{\gamma_{x}}{2}u^{(0)}_{y}+\frac{\gamma_{y}}{2}u^{(0)}_{x},\nonumber\\
\dot{u}^{(1)}_{I}=&\frac{\gamma_{x}}{2}u^{(1)}_{x}+\frac{\gamma_{y}}{2}u^{(1)}_{y}-\frac{u^{(0)}_{z}}{2},~~\dot{u}^{(1)}_{x}=-\frac{\gamma_{y}}{2}u^{(1)}_{z}-\frac{\gamma_{x}}{2}u^{(1)}_{I}-\frac{u^{(0)}_{y}}{2},\nonumber\\
\dot{u}^{(1)}_{y}=&\frac{\gamma_{x}}{2}u^{(1)}_{z}-\frac{\gamma_{y}}{2}u^{(1)}_{I}+\frac{u^{(0)}_{x}}{2},~~\dot{u}^{(1)}_{z}=-\frac{\gamma_{x}}{2}u^{(1)}_{y}+\frac{\gamma_{y}}{2}u^{(1)}_{x}+\frac{u^{(0)}_{I}}{2},\nonumber\\
\dot{\gamma}_{I}=&0,~~\dot{\gamma}_{x}=-\gamma_{y}\gamma_{z}-\delta_{y},~~\dot{\gamma}_{y}=\gamma_{x}\gamma_{z}+\delta_{x},~~\dot{\gamma}_{z}=0,
\dot{\delta}_{I}=0,~~\dot{\delta}_{x}=-\gamma_{y}\delta_{z},~~\dot{\delta}_{y}=\gamma_{x}\delta_{z},~~\dot{\delta}_{z}=\gamma_{y}\delta_{x}-\gamma_{x}\delta_{y},
\label{eq:paoptimal_eq_original}
\end{align}
where dots over the variables represent the time derivative.
It appears a tough task to solve these equations.
We, however, note that the rhs of the equations are Lipschitz continuous, and the solution is unique if exists.
Hence, it is sufficient to anyhow find a solution.

Motivated by the short-CORPSE, which is the shortest known detuning-robust control, we assume that the shortest operation is (a)  a coaxial rotation, i.e., $\omega_{y}(t)=\gamma_{y}(t)=0$ throughout the dynamics, and (b) symmetric: $H_{\rm cont}(t)=H_{\rm cont}(T-t)$.
The equations~(\ref{eq:paoptimal_eq_original}) are then rewritten as
\begin{align}
\dot{u}^{(0)}_{I}=&\frac{\gamma_{x}}{2}u^{(0)}_{x},~~\dot{u}^{(0)}_{x}=-\frac{\gamma_{x}}{2}u^{(0)}_{I},~~
u^{(0)}_{y}=0,~~u^{(0)}_{z}=0,\nonumber\\
u^{(1)}_{I}=&0,~~u^{(1)}_{x}=0,~~\dot{u}^{(1)}_{y}=\frac{\gamma_{x}}{2}u^{(1)}_{z}+\frac{u^{(0)}_{x}}{2},~~\dot{u}^{(1)}_{z}=-\frac{\gamma_{x}}{2}u^{(1)}_{y}+\frac{u^{(0)}_{I}}{2},\nonumber\\
\dot{\gamma}_{I}=&0,~~\dot{\gamma}_{x}=-\delta_{y},~~\dot{\gamma}_{y}=\gamma_{x}\gamma_{z}+\delta_{x}=0,~~\dot{\gamma}_{z}=0,\nonumber\\
\dot{\delta}_{I}=&0,~~\dot{\delta}_{x}=0,~~\dot{\delta}_{y}=\gamma_{x}\delta_{z},~~\dot{\delta}_{z}=-\gamma_{x}\delta_{y}.
\end{align}
The equations for the variables $(u^{(0)}_{y,z},u^{(1)}_{I,x})$ have already been integrated because their equations can be solved independently of the other parts.
The linearlity of the equations and the initial conditions,
\begin{equation}
\left(u^{(0)}_{y}(0),u^{(0)}_{z}(0),u^{(1)}_{I}(0),u^{(1)}_{x}(0)\right)=(0,0,0,0),
\end{equation}
results in their constantness.
In particular, the equality $u^{(0)}_{y}=u^{(0)}_{z}=0$ is physically obvious because we now assume that the operation (without the detuning error) is coaxial.

As easily seen, the non-trivial independent equations are
\begin{equation}
\dot{u}^{(0)}_{I}=\frac{\gamma_{x}}{2}u^{(0)}_{x},~~\dot{u}^{(0)}_{x}=-\frac{\gamma_{x}}{2}u^{(0)}_{I},~~\dot{\gamma}_{x}=-\delta_{y},~~\dot{\delta}_{y}=\gamma_{x}\delta_{z},~~\dot{\delta}_{z}=-\gamma_{x}\delta_{y},~~\dot{u}^{(1)}_{y}=\frac{\gamma_{x}}{2}u^{(1)}_{z}+\frac{u^{(0)}_{x}}{2},~~\dot{u}^{(1)}_{z}=-\frac{\gamma_{x}}{2}u^{(1)}_{y}+\frac{u^{(0)}_{I}}{2},
\label{eq:dynamics_in_pa}
\end{equation}
and the other variables are taken to be zero to guarantee consistency.
We introduce new variables $\theta$ and $\Theta$ such that $u^{(0)}_{I}=\cos(\theta/2)$, $u^{(0)}_{x}=\sin(\theta/2)$, $\delta_{y}=-D\sin\Theta$, and $\delta_{z}=D\cos\Theta$, where $D$ is a constant.
As we now consider a coaxial rotation, $\theta(t)$ is directly identified with the rotation angle of $U^{(0)}(t)$ throughout the dynamics.
Furthermore, it is convenient to introduce 
\begin{equation}
\tilde{u}^{(1)}_{y}=u^{(0)}_{I}u^{(1)}_{y}+u^{(0)}_{x}u^{(1)}_{z},~~
\tilde{u}^{(1)}_{z}=u^{(0)}_{I}u^{(1)}_{z}-u^{(0)}_{x}u^{(1)}_{y}.
\end{equation}
These variables follow from the frame transformation, $U^{(1)}\rightarrow \left(U^{(0)}\right)^{\dagger}U^{(1)}$.
We then obtain
\begin{equation}
\dot{\theta}=-\gamma_{x},~~\dot{\Theta}=-\gamma_{x},~~\dot{\gamma}_{x}=D\sin\Theta,~~\dot{\tilde{u}}^{(1)}_{y}=\frac{\sin\theta}{2},~~\dot{\tilde{u}}^{(1)}_{z}=\frac{\cos\theta}{2},
\end{equation}
where first (second) equation corresponds to the first and second (third and fourth) equations in Eq.~(\ref{eq:dynamics_in_pa}).
These equations are summarized in terms of $\Theta$ as follows:
\begin{equation}
\theta=\lambda+\Theta,~~\ddot{\Theta}+D\sin\Theta=0,~~\int^{T}_{0}d t \cos\Theta=\int^{T}_{0}d t\sin\Theta=0,
\end{equation}
where $\lambda$ is a constant.
The latter integral equations come from the boundary condition $U^{(1)}(T)=0$ in Eq.~(\ref{eq:robust}), i.e., they represent the detuning robustness.
We need to solve the first two equations whilst guaranteeing the last two conditions.
The second one is nothing but the equation for a simple pendulum without the approximation of sufficiently small $\Theta$.
These equations are also derived in a simpler way using the Lagrange multiplier method (Appendix \ref{sec:elmethod}).

The general solution for $\Theta$ is
\begin{equation}
\Theta(t)=2 \am(b t+c,k),
\label{eq:general_eq}
\end{equation}
where $\am$ is the Jacobi's amplitude function and $b$, $c$, and $k$ are determined by the boundary conditions and constraints.
Detailed derivations are shown in Appendix \ref{sec:jacobi}.
Let us now exploit the ansatz (b) of the symmetry, which implies
\begin{equation}
\theta(t)+\theta(T-t)=\theta_{f},~~\Theta(t)=-\Theta(T-t).
\label{eq:symmetric}
\end{equation}
We obtain $\lambda=\theta_{f}/2$ by substituting $t=T/2$ in the first equation and then $\theta(t)=\Theta(t)+\theta_{f}/2$, and $\Theta(0)=-\Theta(T/2)=-\theta_{f}/2$.
Together with Eq.~(\ref{eq:general_eq}), one find
\begin{equation}
\Theta(t)=2 \am \left(\frac{t-T/2}{2},k\right),
\label{eq:solution2}
\end{equation}
where $T$ satisfies $2\am (\pm T,k)=\pm \theta_{f}/2$.
The denominator $2$ is determined so that the maximal speed $\dot{\Theta}(t)$ during the dynamics is fixed to be $1$.
When $k\leq1$, $\am(x,k)$ is a monotonic function and its inverse is uniquely determined by the incomplete elliptic integral of the first kind $F$: $\am\left(F(a,k),k\right)=a$.
Hence, $T=4 F(\theta_{f}/4,k)$.
When $k>1$, in addition to the above $T$, one find other possibilities of $T$.
In this case,  $\am(x,k)$ satisfies $\am(x,k)=\am(2n{\rm Re}\left(K(k)\right)-x,k)$, where $K(k)$ is the complete elliptic integral of the first kind, for any integer $n$ because of its periodicity.
Therefore, one can take $T=4\left(2n{\rm Re}\left(K(k)\right)- F(\theta_{f}/4,k)\right)$.
The possibilities of $n\geq2$ are redundant for our purpose, we only need to consider the following two $T$'s:
\begin{subequations}
\begin{align}
T=&4 F(\theta_{f}/4,k),~{\rm for~any}~k,~~{\rm or}\label{eq:solution:branch3}\\
T=&4\left(2{\rm Re}\left(K(k)\right)- F(\theta_{f}/4,k)\right),~{\rm for}~k>1,
\label{eq:solution:branch4}
\end{align}
\label{eq:solution3}
\end{subequations}
$T$ and $k$ of the solution are chosen so that the constraint
\begin{equation}
\int^{T}_{0}dt \cos \Theta=0,
\label{eq:cosconst2}
\end{equation}
holds, and of course, the choice depends on $\theta_{f}$.
Because of anti-symmetry $\Theta(t)=-\Theta(T-t)$, The other constraint with $\sin \Theta $ is trivially satisfied:
\begin{equation}
\int^{T}_{0}d t \sin\left(\Theta(t)\right)=\int^{T/2}_{0}d t \sin\left(\Theta(t)\right)+\int^{T}_{T/2}d t \sin\left(\Theta(t)\right)=\int^{T/2}_{0}d t \sin\left(\Theta(t)\right)-\int^{T/2}_{0}d t \sin\left(\Theta(t)\right)=0.
\label{eq:symmetry}
\end{equation}
If we find appropriate $T$ and $k$ for a certain $\theta_{f}$, then Eq.~(\ref{eq:solution2}) is the unique solution of Eq.~(\ref{eq:paoptimal_eq_original}) up to trivial ambiguities such as periodicity.

\begin{figure}[h]
		\begin{center}
			\includegraphics[width=130mm]{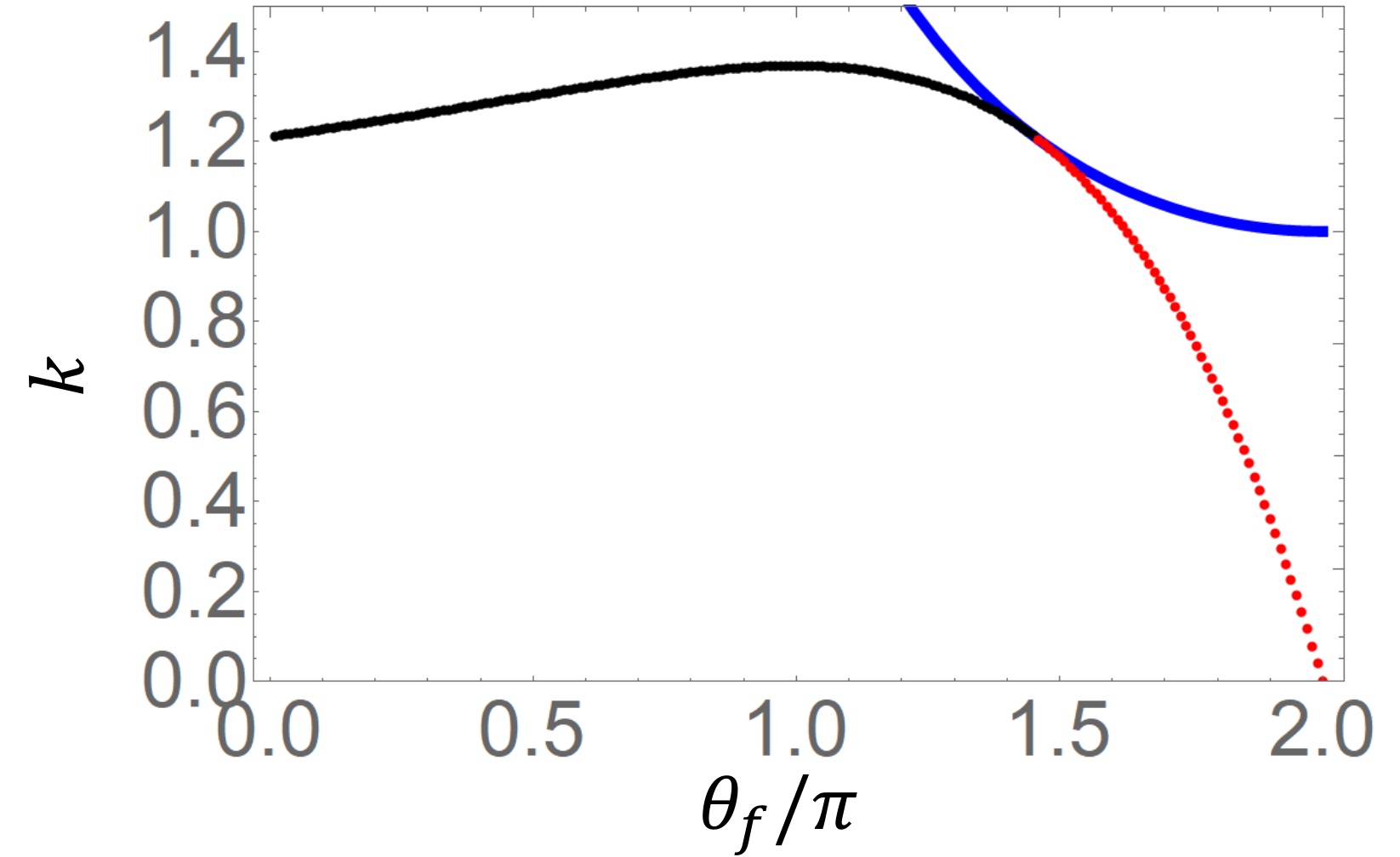}
			\caption{The parameter $k$ for satisfying the detuning robustness as a function of $\theta_{f}$.
			The black and red dots represent $k$ while the blue line shows the upper limit of $k$: $k_{\sup}:=1/\sin^{2}(\theta_{f}/4)$.
			For the parameter $k$ depicted by the black dots, the operation time $T$ is taken from Eq.~(\ref{eq:solution:branch4}), whilst  the region of the red dots has the operation time from Eq.~(\ref{eq:solution:branch3}).
			\label{fig:1}
			}
		\end{center}
	\end{figure}

Figure~\ref{fig:1} shows the appropriate parameter $k$ found for each $\theta_{f}$.
It appears that for any $\theta_{f}\in [0,2\pi]$, there exists the parameter $k$ that satisfies Eq.~(\ref{eq:cosconst2}).
The parameter $k$ is $0$ when $\theta_{f}=2\pi$, which implies that the direct $2\pi$ rotation with a constant speed is detuning-robust.
This fact is well-known in common treatment of composite pulses in NMR.
Once the parameter $k$ touches the line $k_{\sup}:=1/\sin^{2}(\theta_{f}/4)$, which is the upper limit of $k$ given by the condition that $F(\theta_{f}/4,k)$ is real, $T$ of the solution changes from Eq. (\ref{eq:solution:branch4}) to (\ref{eq:solution:branch3}).
We should carefully treat the case of $\theta_{f}=0$, where we actually have two possible solutions.
One is a natural extrapolation in Fig.~\ref{fig:1}, which gives $k\sim 1.2$, and this is one lap of a non-trivial oscillation with a period $T$.
On the other hand, we obtain the identity dynamics with $T=0$.
The latter is obviously the true optimum.
Note that this ambiguity arises from the periodicity of the solution, and does not violate its (non-trivial) uniqueness.

\subsection{time optimization}

The original equations in the $L^{(t)}$ optimization is similar to Eqs.~(\ref{eq:paoptimal_eq_original}):
\begin{align}
\dot{u}^{(0)}_{I}=&\frac{\bar{\gamma}_{x}}{2}u^{(0)}_{x}+\frac{\bar{\gamma}_{y}}{2}u^{(0)}_{y},~~\dot{u}^{(0)}_{x}=-\frac{\bar{\gamma}_{y}}{2}u^{(0)}_{z}-\frac{\bar{\gamma}_{x}}{2}u^{(0)}_{I},\nonumber\\
\dot{u}^{(0)}_{y}=&\frac{\bar{\gamma}_{x}}{2}u^{(0)}_{z}-\frac{\bar{\gamma}_{y}}{2}u^{(0)}_{I},~~\dot{u}^{(0)}_{z}=-\frac{\bar{\gamma}_{x}}{2}u^{(0)}_{y}+\frac{\bar{\gamma}_{y}}{2}u^{(0)}_{x},\nonumber\\
\dot{u}^{(1)}_{I}=&\frac{\bar{\gamma}_{x}}{2}u^{(1)}_{x}+\frac{\bar{\gamma}_{y}}{2}u^{(1)}_{y}-\frac{u^{(0)}_{z}}{2},~~\dot{u}^{(1)}_{x}=-\frac{\bar{\gamma}_{y}}{2}u^{(1)}_{z}-\frac{\bar{\gamma}_{x}}{2}u^{(1)}_{I}+\frac{u^{(0)}_{y}}{2},\nonumber\\
\dot{u}^{(1)}_{y}=&\frac{\bar{\gamma}_{x}}{2}u^{(1)}_{z}-\frac{\bar{\gamma}_{y}}{2}u^{(1)}_{I}-\frac{u^{(0)}_{x}}{2},~~\dot{u}^{(1)}_{z}=-\frac{\bar{\gamma}_{x}}{2}u^{(1)}_{y}+\frac{\bar{\gamma}_{y}}{2}u^{(1)}_{x}-\frac{u^{(0)}_{I}}{2},\nonumber\\
\dot{\gamma}_{I}=&0,~~\dot{\gamma}_{x}=-\bar{\gamma}_{y}\gamma_{z}-\delta_{y},~~\dot{\gamma}_{y}=\bar{\gamma}_{x}\gamma_{z}+\delta_{x},~~\dot{\gamma}_{z}=0,\nonumber\\
\dot{\delta}_{I}=&0,~~\dot{\delta}_{x}=-\bar{\gamma}_{y}\delta_{z},~~\dot{\delta}_{y}=\bar{\gamma}_{x}\delta_{z},~~\dot{\delta}_{z}=\bar{\gamma}_{y}\delta_{x}-\bar{\gamma}_{x}\delta_{y},
\label{eq:timeoptimal_eq_original}
\end{align}
where $\gamma_{x,y}$ in the pulse-area optimization are replaced by $\bar{\gamma}_{x,y}$ (cf. Eqs.~(\ref{eq:time}) and (\ref{eq:optimal_pa})).
Similarly, we follow the ans\"aze (a) and (b), and finally obtain
\begin{equation}
\theta=\lambda+\Theta,~~\dot{\Theta}={\rm sgn}\left(\gamma_{x}\right),~~\dot{\gamma}_{x}=D \sin\Theta,~~\int^{T}_{0}d t \cos\Theta=\int^{T}_{0}d t\sin\Theta=0,
\label{eq:time-optimal}
\end{equation}
where we use the equality ${\rm sgn}\left(\gamma_{x}\right)=\gamma_{x}/\left|\gamma_{x}\right|$.
The solutions of these equations are not trivial.
One can, however, see that the short-CORPSE in Eq.~(\ref{eq:shortCORPSE}) with $\Omega=1$ satisfies them by taking $\lambda=\theta_{f}/2$ and the following auxiliary variables:
\begin{align}
\Theta(t)=&
\begin{cases}
-  t-\theta_{f}/2,&0\leq  t \leq \theta^{\rm (SC)}_{1},\\
  t-2\theta^{\rm (SC)}_{1}-\theta_{f}/2,&\theta^{\rm (SC)}_{1} \leq  t \leq \theta^{\rm (SC)}_{1}+\theta^{\rm (SC)}_{2},\\
-  t+\theta^{\rm (SC)}_{2}-\theta_{f}/2,&\theta^{\rm (SC)}_{1}+\theta^{\rm (SC)}_{2}\leq  t \leq 2 \theta^{\rm (SC)}_{1}+\theta^{\rm (SC)}_{2},
\end{cases}
\nonumber\\
\gamma_{x}(t)=&
\begin{cases}
-\cos\Theta(t)-\cos\kappa,&0\leq  t \leq \theta_{1}^{\rm (SC)},\\
\cos\Theta(t)+\cos\kappa,& \theta_{1}^{\rm (SC)}\leq t \leq  \theta_{1}^{\rm (SC)}+ \theta_{2}^{\rm (SC)},\\
-\cos\Theta(t)-\cos\kappa,& \theta_{1}^{\rm (SC)}+ \theta_{2}^{\rm (SC)}\leq t \leq  2 \theta_{1}^{\rm (SC)}+ \theta_{2}^{\rm (SC)}.
\end{cases}
\end{align}
We consider other solutions with drawing the phase portrait in Appendix~\ref{app:corpse}.
The short-CORPSE is the optimal control among these solutions.
However, we should note that in the case of the time optimization, one cannot conclude the solution obtained with the ans\"atze (short-CORPSE) is truly optimal although it is expected to be the shortest:
the original equations are not Lipschitz continuous (discontinuous at $(\gamma_{x},\gamma_{y})=(0,0)$), and then the uniqueness of the solution are not guaranteed.

\section{behaviours of optimal solutions}
\label{sec:IIII}

\begin{figure}[h]
		\begin{center}
			\includegraphics[width=170mm]{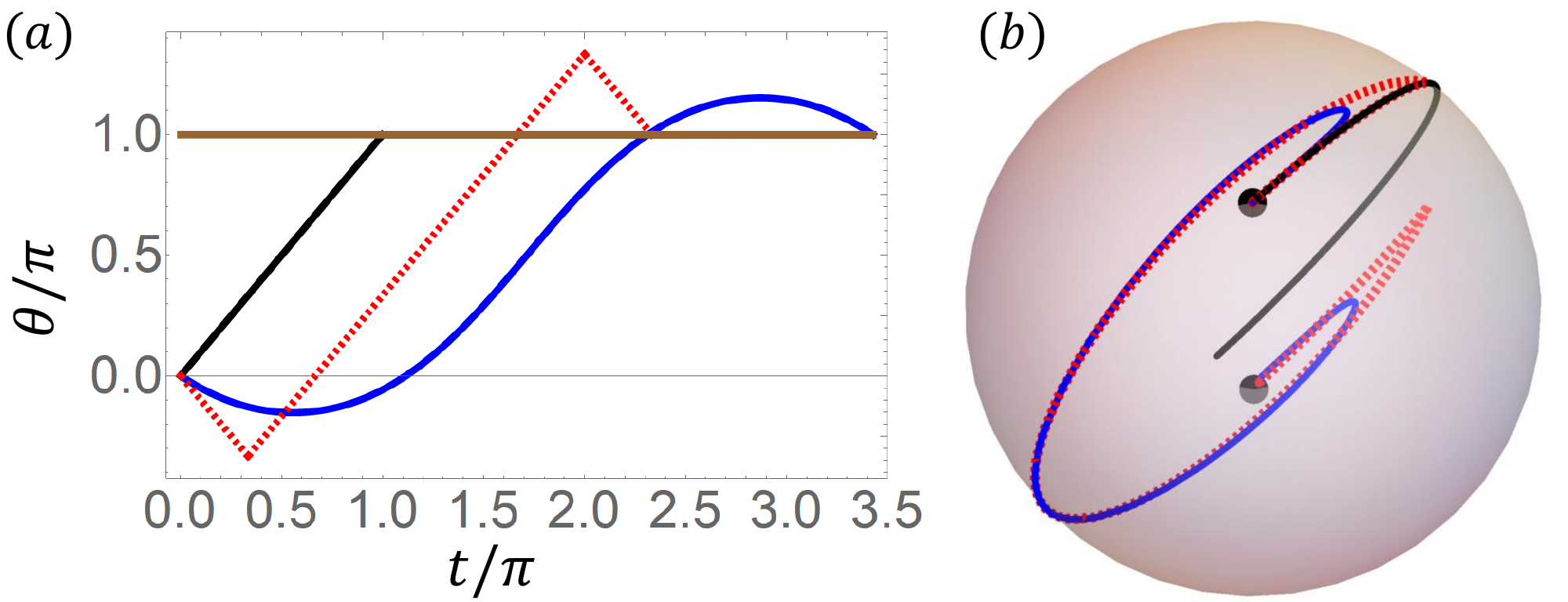}
			\caption{The trajectory $\theta(t)$ for $\theta_{f}=\pi$.
			(a) $\theta(t)/\pi$ as a function of $t$.
			The blue line is $\theta(t)$, the red dashed line is the short-CORPSE, and the brown one represents $\theta_{f}=\pi$.
			(b) Bloch sphere representation of the trajectory with the detuining error.
			The trajectory of the pulse-area optimal operation is depicted by blue while the direct one is in black.
			The short-CORPSE trajectory is represented by the red dashed line.
			The error strength is taken to be $f=0.1$.
			\label{fig:2}
			}
		\end{center}
	\end{figure}

We have obtained the pulse-area and time (probably) optimal solutions with the detuning robustness.
In this section, we show their behaviors and properties as a function of the target rotation angle $\theta_{f}$.
Let us first show that their trajectories with and without the detuning error for several values of $\theta_{f}$.
In Fig.~\ref{fig:2}(a), we provide the time and pulse-area optimal dynamics of $\theta(t)$ for the target operation $\bar{R}(\theta_{f}=\pi,0)$.
The dynamics of $\theta(t)$ corresponds to the rotation angle $\theta$ of the errorless dynamics $U^{(0)}(t)$ at each moment (see Eq.~(\ref{eq:rot_rep})) because our solution is a coaxial rotation~($\vec{k}={\rm const}$).
Meanwhile, Fig.~\ref{fig:2}(b) exhibit the actual trajectory of this dynamics under the detuning error with the strength $f=0.1$.
An important characteristic is that both trajectories have ``switchback" behaviors at some appropriate timings.
These switchbacks work to compensate for the detuning error.
The pulse-area optimal trajectory is smooth, and not always with its maximal speed $1(=\Omega)$ whereas the short-CORPSE always keeps the maximal speed and the changes are abrupt.

\begin{figure}[h]
		\begin{center}
			\includegraphics[width=170mm]{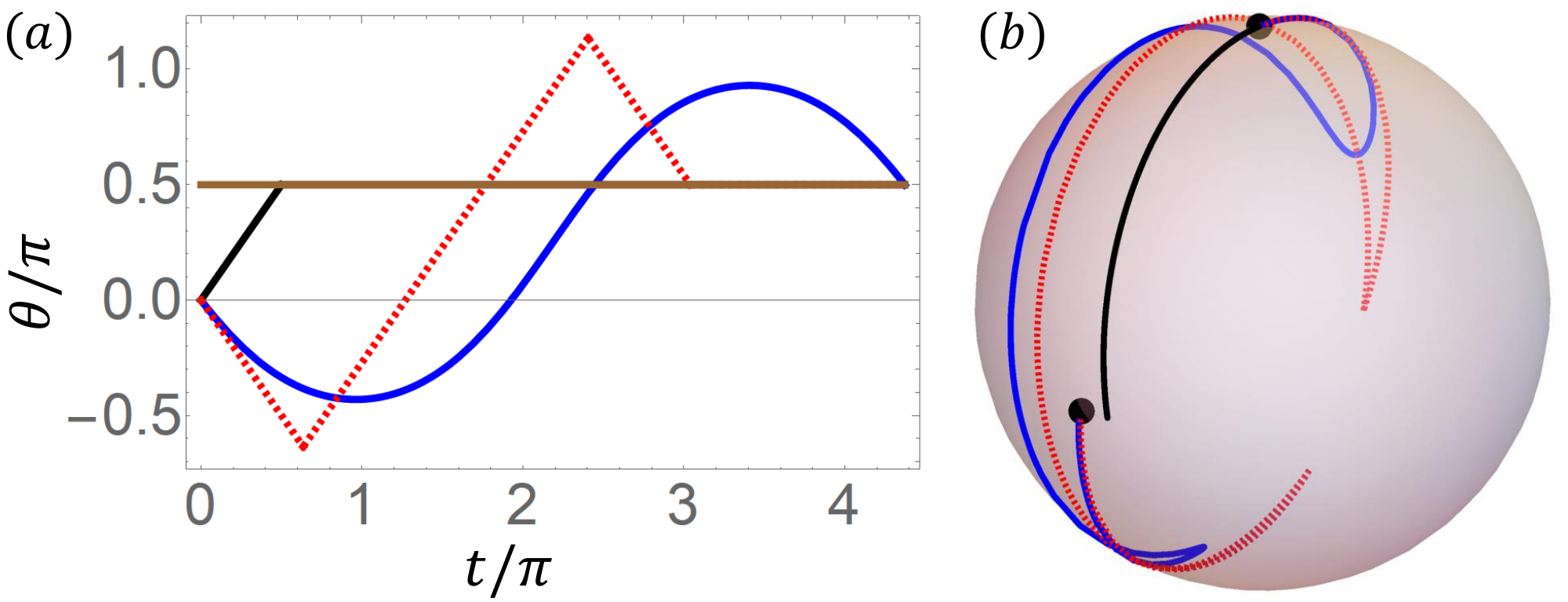}
			\caption{The trajectory $\theta(t)$ for $\theta_{f}=\pi/2$.
							(a) $\theta(t)/\pi$ as a function of $t$.
							The blue line is $\theta(t)$, the red dashed line is the short-CORPSE, and the brown one represents $\theta_{f}=\pi/2$.
							(b) Bloch sphere representation of the trajectory with the detuning error.
							The trajectory of the pulse-area optimal operation is depicted by blue while the direct one is in black.
							The short-CORPSE trajectory is represented by the red dashed line.
							The error strength is taken to be $f=0.1$.
			\label{fig:3}
			}
		\end{center}
	\end{figure}

Figure~\ref{fig:3} shows the two optimal controls for $\theta_{f}=\pi/2$ in the same way as Fig.~\ref{fig:2}.
As one can see from Fig.~\ref{fig:3}, both optimal controls have relatively dynamic switchback behavior compared with the controls for $\theta_{f}=\pi$.
This is related to an intuitive explanation of the detuning robustness introduced in Ref.~\cite{kukita2022geometric}.
For a detuning-robust control, its ``mass center" of the trajectory in the Bloch sphere representation must lie on the xy plane.
In particular, coaxial robust controls must have their mass center at the origin.
As the trajectory of the direct operation for $\theta_{f}=\pi/2$ does not step on the south hemisphere, a drastic deformation from the direct operation is required to shift its mass center to the origin, i.e., to attain the detuning robustness.
Meanwhile, the trajectory of the direct operation for $\theta_{f}=\pi$ passes through the south hemisphere, and thus a relatively slight deformation is sufficient.

\begin{figure}[h]
		\begin{center}
			\includegraphics[width=170mm]{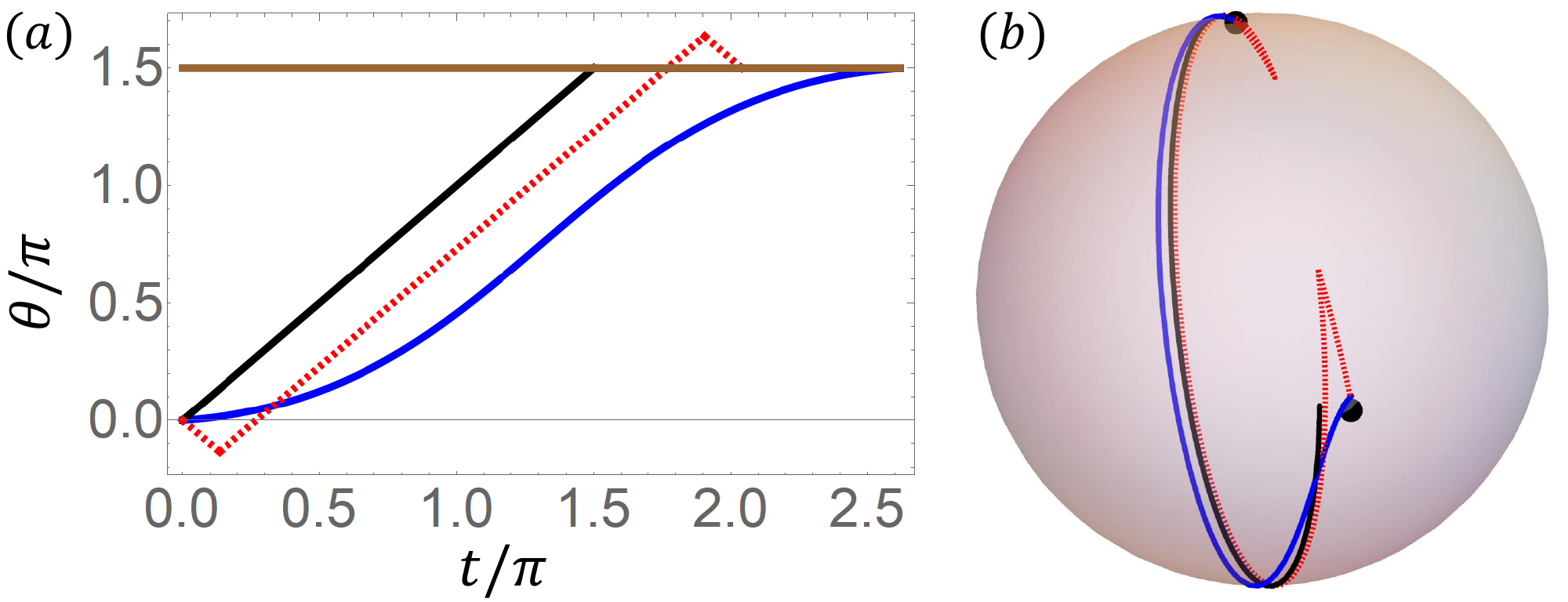}
			\caption{The trajectory $\theta(t)$ for $\theta_{f}=3\pi/2$.
						(a) $\theta(t)/\pi$ as a function of $t$.
						The blue line is $\theta(t)$, the red dashed line is the short-CORPSE, and the brown one represents $\theta_{f}=3\pi/2$.
						(b) Bloch sphere representation of the trajectory with the detuning error.
						The trajectory of the pulse-area optimal operation is depicted by blue while the direct one is in black.
						The short-CORPSE trajectory is represented by the red dashed line.
						The error strength is taken to be $f=0.1$.
			\label{fig:4}
			}
		\end{center}
	\end{figure}

In Fig.~\ref{fig:4}, the optimal controls for $\theta_{f}=3\pi/2$ are depicted.
Notably, the pulse-area optimal control is monotonic with respect to time whereas the short-CORPSE still has switchbacks.
This behavior is not the same as the case of $\theta_{f}=\pi/2$ even though the target operation for these cases are the same up to the global phase of the unitary matrix.
The discord between both cases is intuitively explained by the aforementioned mass-center representation of the detuning robustness \cite{kukita2022geometric}:
As the trajectory of the direct operation for $\theta_{f}=3\pi/2$ has the mass center near the origin comparing with the case of $\theta_{f}=\pi/2$, a slight modification of the dynamics is sufficient to attain the robustness.
As discussed later, the monotonicity of the pulse-area optimal control makes its pulse area be the same as that of the direct operation.
This is true for all $\theta_{f}\gtrapprox 1.46$, which corrsponds to the solution with $T$ of Eq.~(\ref{eq:solution:branch3}) corresponding to the red dots in Fig.~\ref{fig:1}.
In this parameter region, controlling the speed without changing its sign is sufficient to compensate for the detuining error.
The curved pulse-area optimal control gradually approaches to a straight line as $\theta_{f}$ increases, and finally become straight, i.e., the direct operation when $\theta_{f}=2\pi$.

Although the short-CORPSE has switchback behaviors even in this case, it is relatively small compared with the case of $\theta_{f}=\pi$.
As increasing $\theta_{f}$, we observe that these switchback behaviors get smaller, and the control finally corresponds to the direct operation when $\theta_{f}=2\pi$, as well as the pulse-area optimal one.

\begin{figure}[h]
		\begin{center}
			\includegraphics[width=180mm]{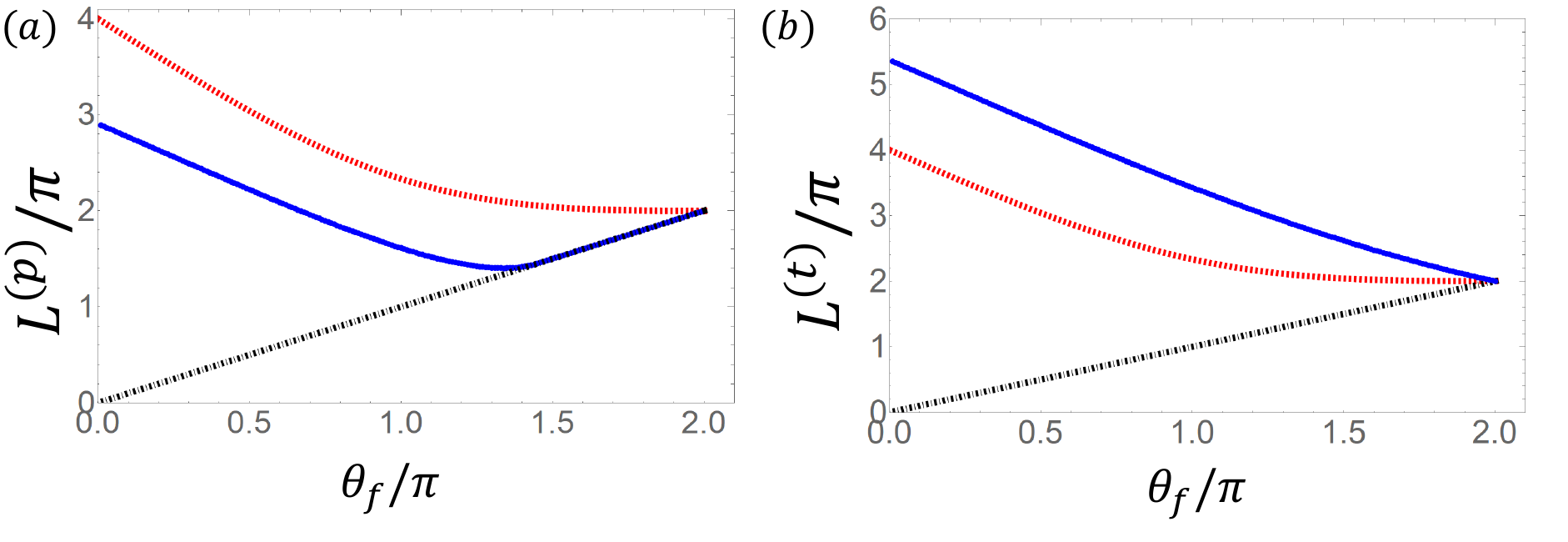}
			\caption{
			(a) The pulse area $L^{(p)}$ and (b) the operation time $L^{(t)}=T$ of operations.
			The blue dots are the pulse area (operation time) of the found pulse-area optimal operation for $\theta_{f}\in [0,2\pi]$.
			The black dot-dashed line represents the pulse area (operation time) of the direct operation (=$\theta_{f}$).
			The red line is the pulse area (operation time) of the short CORPSE sequence.
			\label{fig:5}
			}
		\end{center}
	\end{figure}

Figure~\ref{fig:5} shows the pulse area and the operation time as a function of $\theta_{f}$.
In Fig.~\ref{fig:5}(a), the pulse areas of the short-CORPSE and the pulse-area optimal control found in this paper are depicted.
As aforementioned, the pulse-area optimal control is monotonic when $\theta_{f}\gtrapprox 1.46$, and thus its pulse area is the same as that of the direct operation.
Compared with the short-CORPSE with abrupt changes in its speed, the smoothness of the pulse-area optimal control offer an advantage for saving its pulse area.

Figure~\ref{fig:5}(b) exhibits the operation time $L^{(t)}(=T)$ of the optimal controls.
For both cases, $L^{(t)}$ monotonically decreases as $\theta_{f}$ increases.
The short-CORPSE has the same operation time as its pulse-area because the control speed is always $1(=\Omega)$ throughout the dynamics.
Meanwhile, $L^{(t)}$ and $L^{(e)}$ of the pulse-area optimal control do not coincide: its speed smoothly changes during the dynamics.
Although $L^{(t)}$ of the short-CORPSE beats that of the pulse-area optimal one, this does not mean that the short-CORPSE is truly time optimal.
Our ans\"atze are too restrictive to find the truly time optimal control unlike the case of the pulse-area optimality.

\section{summary}
\label{sec:IIIII}

In this paper, we have obtained the pulse-area and (probably) time optimal control robust against detuning error using the Pontryagin's Maximum Principle (PMP).
Our target operation is in the form of a rotation with the angle of $\theta_{f}$ in the Bloch sphere representation, but its axis is restricted on the $xy$ plane.
This form includes important operations, such as the NOT gate, and the generation of superposition states.
Unlike Ref.~\cite{PhysRevA.95.063403}, we construct optimal controls without initial-state dependence and thus their applicability would be wider.

We found that a smooth trajectory of $\theta(t)$ described by the Jacobi's amplitude function is the pulse-area optimal solution.
As discussed in this paper, the Lipschitz continuity of the equations to be solved guarantees the uniqueness of the solution even though we imposed several ans\"atze in the derivation.
The solution has two possible $T$'s depending on $\theta_{f}$: with and without ``switchbacks".
For sufficiently large $\theta_{f}\gtrapprox 1.46 \pi$, the solution trajectory in the Bloch sphere has no switchback behaviors, and thus its pulse area corresponds to that of the direct operation.

We have also shown that  the short-CORPSE, which is the shortest known detuning-robust control, is a candidate of the time optimal control.
We cannot guarantee that this is the truly optimal solution because of the lack of the Lipschitz continuity in the equations.
However, as this kind of coaxial rotations is simply implemented experimentally, the short-CORPSE will effectively be appropriate in realistic situations.
Further theoretical investigation will be required.

Optimal controls under the first-order detuning robustness have only been considered in this paper.
The optimality of higher-order detuning-robust controls would be important in practice, which can be our future work.
As higher-order robustness will require longer operation time and larger pulse area, we should more carefully evaluate their effectiveness with taking realistic effect of decoherence into account.

\appendix

\section{Pontryagin's Maximum Principle}
\label{sec:pmp}

In this appendix, we introduce the Pontryagin's maximum principle (PMP) \cite{pontryagin2018mathematical}.
Consider the following differential equations:
\begin{equation}
\frac{d x^{i} (t)}{d t}=f^{i}\left(\vec{x}(t);\vec{u}(t)\right),~~\vec{x}(t):=\left(x^{1}(t),\cdots,x^{n}(t)\right),
~\vec{u}(t):=\left(u^{1}(t),\cdots,u^{m}(t)\right),
\label{eq:origin}
\end{equation}
where $x^{i}(t)$'s are dynamical variables while $u^{j}(t)$'s control parameters.
Our goal is to minimize (or maximize) a cost function, such as total energy consumption during the dynamics, and the time required for a target operation, by adjusting the control parameters $\vec{u}(t)$.
When $\vec{u}$ is restricted in a finite region, the ordinary variational principle is no longer valid:
one cannot take the variation of the cost function with respect to $\vec{u}$ on boundaries of the region.

Pontryagin proposed a optimization method under inequality constraints describing that the control parameters are contained in the finite region $\Omega$.
Consider Eq.~(\ref{eq:origin}) with an initial conditions $\vec{x}(0)=\vec{x}_{I}$ at $t=0$ and a final condition $\vec{x}(t_{f})=\vec{x}_{F}$ at $t=t_{f}$.
In general, there are many possibilities in scheduling of $\vec{u}(t)$ satisfying these initial and final conditions.
Our task is to find a scheduling of $\vec{u}(t)$ for minimizing (or maximizing) a cost function in the following form:
\begin{equation}
L=\int^{t_{f}}_{0}d t f^{0}\left(\vec{x}(t);\vec{u}(t)\right).
\label{eq:cost}
\end{equation}
Typically, the cost function represents the total consumption of some quantities, such as energy, with the fixed initial and final conditions. 
In particular, the optimization of the time required for the dynamics is represented by $f^{0}\left(\vec{x}(t);\vec{u}(t)\right)=1$.
To solve this problem, we define an auxiliary variable,
\begin{equation}
x^{0}(t)=\int^{t}_{0}d t f^{0}\left(\vec{x}(t);\vec{u}(t)\right),
\end{equation}
which satisfies the equation,
\begin{equation}
\frac{d x^{0}(t)}{d t}= f^{0}\left(\vec{x}(t);\vec{u}(t)\right).
\end{equation}
Thus, the initial and final condition of the extended variable vector $\vec{\chi}(t)=\left(x^{0}(t),x^{1}(t),\cdots,x^{n}(t)\right)$ are given by
\begin{equation}
\vec{\chi}(0)=\left(0,\vec{x}_{I}\right),~~
\vec{\chi}(t_{f})=\left(L,\vec{x}_{F}\right).
\label{eq:boundary}
\end{equation}

We also introduce the ``conjugate momenta" $\vec{p}(t)=\left(p_{0}(t),p_{1}(t),\cdots,p_{n}(t)\right)$ through the equations,
\begin{equation}
\frac{d p_{i}(t)}{d t}=-\sum^{n}_{j=0}\frac{\partial f^{j}\left(\vec{x}(t);\vec{u}(t)\right)}{\partial x^{i}}p_{j}(t),
\label{eq:cano_for_p}
\end{equation}
which are linear and homogeneous with respect to $\vec{p}(t)$.
Defining the ``Hamiltonian",
\begin{equation}
H(\vec{p},\vec{x};\vec{u})=\sum_{n}^{i=0}p_{i}f^{i}\left(\vec{x};\vec{u}\right),
\end{equation}
we reformulate the dynamics of $\vec{p}(t)$ and $\vec{\chi}(t)$ in the form of Hamilton's equations,
\begin{equation}
\frac{d x^{i}}{d t}=\frac{\partial H\left(\vec{p}(t),\vec{x}(t);\vec{u}(t)\right)}{\partial p_{i}},~~\frac{d p_{i}}{d t}=-\frac{\partial H\left(\vec{p}(t),\vec{x}(t);\vec{u}(t)\right)}{\partial x^{i}},~~i=0,1,\cdots,n.
\label{eq:canon}
\end{equation}
In particular, the $0$-th momentum $p_{0}(t)$ is constant during the dynamics:
\begin{equation}
\frac{d p_{0}(t)}{d t}=0,
\end{equation}
because $H\left(\vec{p},\vec{x};\vec{u}\right)$ does not contain $x^{0}$.

The PMP is summarized as follows:
For a control $\vec{u}^{*}(t)$ minimizing the cost function $L$ and the corresponding trajectory $\vec{x}^{*}(t)$ satisfying the equation (\ref{eq:origin}) and the boundary conditions (\ref{eq:boundary}), then there exists a non-zero vector $\vec{p}^{*}(t)$ that satisfies the equation (\ref{eq:canon}) and,
\begin{equation}
H\left(\vec{p}^{*}(t),\vec{x}^{*}(t);\vec{u}^{*}(t)\right)=K\left(\vec{p}^{*}(t),\vec{x}^{*}(t)\right):=\max_{\vec{v}\in\Omega}H\left(\vec{p}^{*}(t),\vec{x}^{*}(t);\vec{v}\right),~~t_{i}\leq t \leq t_{f}.
\label{eq:optimo}
\end{equation}
In particular, if one has such a vector $\vec{p}(t)$,
\begin{equation}
H\left(\vec{p}^{*}(t),\vec{x}^{*}(t);\vec{u}^{*}(t)\right)=0,
\label{eq:null}
\end{equation}
is satisfied during the dynamics.
Note that this is a necessary condition for the optimization;
in general, we can find several solutions, one of which really optimizes the cost function.

Practically, we first solve Eq. (\ref{eq:optimo}) with respect to $\vec{u}$, i.e.,
\begin{equation}
\vec{u}^{*}(\vec{p},\vec{x})=\underset{\vec{v}\in \Omega}{\rm argmax}~H(\vec{p},\vec{x};\vec{v}).
\end{equation}
Then, solving Eq. (\ref{eq:canon}) with this control $\vec{u}^{*}$,
\begin{equation}
\frac{d x^{i}}{d t}=\frac{\partial H\left(\vec{p}(t),\vec{x}(t);\vec{u}^{*}\left(\vec{p}(t),\vec{x}(t)\right)\right)}{\partial p_{i}},~~\frac{d p_{i}}{d t}=-\frac{\partial H\left(\vec{p}(t),\vec{x}(t);\vec{u}^{*}\left(\vec{p}(t),\vec{x}(t)\right)\right)}{\partial x^{i}},~~i=0,1,\cdots,n,
\label{eq:canoptimo}
\end{equation}
we obtain a candidate of the optimal trajectory.
Let us explain how to fix free parameters in a solution.
As Eq. (\ref{eq:canoptimo}) is a set of $2n+2$ first-order differential equations, and $t_{f}$ is also a free parameter, we need to fix $2n+3$ parameters from given conditions.
We, however, have $(2n+1)$~conditions from $\vec{\chi}(0)=(0,\vec{x}_{I})$ and $\vec{x}(t_{f})=\vec{x}_{F}$ ($L$ is determined after the optimization), and thus two degrees of freedom appears to remain.
One of them is absorbed into the absolute value of $\vec{p}$, which is arbitrary by definition because the equations for $\vec{p}$ (\ref{eq:cano_for_p}) are homogeneous.
The other is determined by Eq.~(\ref{eq:null}).
Hence, the number of free parameters matches that of the conditions.

\section{action of optimal controls robust against detuning error}
\label{sec:elmethod}

We reconsider the pulse-area (or energy) optimization via the Lagrange multiplier method, which is simpler than the PMP in the main text.
The reason why we used the PMP is that we have the inequality constraint $|\omega |^{2}=\omega_{x}^{2}+\omega^{2}_{y}\leq \Omega$.
However, when we optimize the pulse area, $|\omega|$ in the cost function works as a penalty term, and thus $\left(\omega_{x},\omega_{y}\right)$ is naturally restricted in a finite region.
This implies that we can solve the pulse-area optimization problem without considering the inequality constraint.

We represent a unitary matrix by the variables $(\theta'=\theta/2,\vartheta,\varphi)$ in the unitary $U$ defined as
\begin{equation}
U=\cos(\theta'){\mathbb I}_{2}-i \sin(\theta')\left(\sin(\vartheta)\cos(\varphi)\sigma_{x}+\sin(\vartheta)\sin(\varphi)\sigma_{y}+\cos(\vartheta)\sigma_{z}\right).
\end{equation}
We take the maximum strength of the control field to be $1$ without loss of generality, and then control this unitary from $t=0$ to $t=T$ with the boundary condition,
\begin{equation}
U(0)={\mathbb I}_{2},~~U(T)=\bar{R}(\theta_{f},0)=\cos(\theta_{f}/2){\mathbb I}_{2}-i \sin(\theta_{f}/2)\sigma_{x},
\label{eq:target}
\end{equation}
which is the rotation with the angle $\theta_{f}$ and the direction $\vec{x}=(1,0,0)$.
Owing to the symmetry of the detuining error,
any detuning-robust operations in the form of $\bar{R}(\theta_{f},\phi_{f})$ can be constructed in the same way as $\bar{R}(\theta_{f},0)$.

The cost function to be minimized is
\begin{equation}
S_{0}(\theta',\vartheta,\varphi)=\int^{T}_{0}dt \tr \left(H^2(t)\right)=\int^{T}_{0}dt \tr \left(\frac{d U(t)}{d t}\frac{d U^{\dagger}(t)}{d t}\right)=\int^{T}_{0}dt \left(\dot{\theta}'^{2}+\sin^{2}\theta'\left(\dot{\vartheta}^{2}+\dot{\varphi}^{2}\sin^{2}\vartheta\right)\right),
\end{equation}
which corresponds to the squared pulse area, or energy consumption $L^{(e)}$ in the main text.
We represent the local constraint that the control field has no $z$ components during the dynamics as
\begin{equation}
\forall t~~~L_{\rm local}(t):=\tr\left(\sigma_{z}H(t)\right)=\dot{\theta}\cos \vartheta-\dot{\vartheta}\cos \theta' \sin\theta'\sin \vartheta+\dot{\varphi}\sin^{2}\theta' \sin^{2}\vartheta=0.
\label{eq:local}
\end{equation}
Also, we impose the global constraint regarding the detuining robustness:
\begin{align}
\int^{T}_{0} dt U^{\dagger}(t)\sigma_{z}U(t)&=0 \longrightarrow\nonumber\\
S_{1}&:=\int^{T}_{0}dt \left(\cos^{2}\theta'+\sin^{2}\theta'\left(\cos^{2}\vartheta-\sin^{2}\vartheta\right)\right)=0\nonumber\\
S_{2}&:=\int^{T}_{0}dt\sin \theta' \cos \theta' \sin \vartheta=0 \nonumber\\
S_{3}&:=\int^{T}_{0}dt \sin^{2}\theta' \cos \vartheta \sin \vartheta=0.
\label{eq:app_constraint}
\end{align}
Thus, the action to be considered is
\begin{equation}
S_{\rm total}=S_{0}+\sum^{3}_{i=1}\lambda_{i}S_{i}+S_{\rm local},~~~S_{\rm local}=\int^{T}_{0}dt \Lambda(t)L_{\rm local}(t),
\end{equation}
where $\Lambda(t)$ is a time-dependent Lagrange multiplier for the local constraint while $\lambda_{1,2,3}$ are constant multipliers for the global constraints.

The whole Euler-Lagrange equations are given as
\begin{align}
\delta_{\theta'}S_{\rm total}=&\frac{d}{d t}\left(2\dot{\theta'}+\Lambda(t)\cos \vartheta \right)-\dot{\vartheta}^{2}\sin\left(2\theta'\right)-\dot{\varphi}^{2}\sin\left(2\theta'\right)\sin^{2}\vartheta
\nonumber\\
&+2\lambda_{1}\sin \left(2\theta'\right)\sin^{2}\vartheta-
\lambda_{2}\cos\left(2\theta'\right)\sin\vartheta-\lambda_{3}\sin\left(2\theta'\right)\cos\vartheta\sin\vartheta\nonumber\\
&+\Lambda(t)\left(\dot{\vartheta}\cos\left(2\theta'\right)\sin\vartheta-\dot{\varphi}\sin\left(2\theta'\right)\sin^{2}\vartheta\right)=0\nonumber\\
\delta_{\vartheta}S_{\rm total}=&\frac{d}{d t}\left(2\dot{\vartheta}\sin^{2}\theta'-\Lambda(t)\cos\theta'\sin\theta'\sin\vartheta\right)-\dot{\varphi}^{2}\sin^{2}\theta'\sin\left(2\vartheta\right)\nonumber\\
&+2\lambda_{1}\sin^{2}\theta'\sin\left(2\vartheta\right)-\lambda_{2}\cos\theta'\sin\theta'\cos\vartheta-\lambda_{3}\sin^{2}\theta'\cos\left(2\vartheta\right)\nonumber\\
&+\Lambda(t)\left(\dot{\theta}'\sin\vartheta+\dot{\vartheta}\cos\theta'\sin\theta'\cos\vartheta-\dot{\varphi}\sin^{2}\theta'\sin\left(2\vartheta\right)\right)=0\nonumber\\
\delta_{\varphi}S_{\rm total}=&\frac{d}{d t}\left(\left(2 \dot{\varphi}+\Lambda(t)\right)\sin^{2}\theta'\sin^{2}\vartheta\right)=0,
\label{eq:lag_original}
\end{align}
and the constraints (\ref{eq:local}) and (\ref{eq:app_constraint}).
The boundary conditions are $\theta(0)=0$, $\theta(T)=\theta_{f}$, $\vartheta(T)=\pi/2$ and $\varphi(T)=0$.
$\vartheta(0)$ and $\varphi(0)$ are arbitrary in general. 
Following the same way in the main text, we make the ans\"{a}tze motivated by the short-CORPSE:
\begin{enumerate}[(a)]
\item The control field $H(t)$ corresponding to the true solution represents a coaxial rotation:~$\vartheta(t)=\frac{\pi}{2}$, $\varphi(t)=0$, $\dot{\vartheta}(t)=\dot{\varphi}(t)=0$, for all $0\leq t \leq T$.
\item The control field is symmetric: $H(t)=H(T-t)$.
\end{enumerate}
According to (a), the original equations~(\ref{eq:lag_original}) are reduced to
\begin{align}
2\ddot{\theta'}=&-2\lambda_{1}\sin \left(2\theta'\right)+\lambda_{2}\cos\left(2\theta'\right)\nonumber\\
\frac{d}{dt}\left(\Lambda(t)\cos\theta'\sin\theta'\right)=&\lambda_{3}\sin^{2}\theta'+\Lambda(t)\dot{\theta}'\nonumber\\
\frac{d}{d t}\left(\Lambda(t)\sin^{2}\theta'\right)=&0\nonumber\\
S_{1}=\int^{T}_{0}d t \cos(2\theta')=&0,\nonumber\\
S_{2}=\frac{1}{2}\int^{T}_{0}d t \sin(2\theta')=&0.
\label{eq:original}
\end{align}
Note the trivial attainment of the constraints $S_{\rm local}=S_{3}=0$.
Moreover, we make additional ans\"{a}tze $\Lambda(t)=\lambda_{3}=0$, which implies that the true solution is a coaxial rotation even when we do not consider $S_{\rm local}$ and $S_{3}$.
The equations are eventually reduced to
\begin{align}
2\ddot{\theta}'&+\lambda_{r}\sin\left(2\theta'+\lambda_{a}\right)=0,~~\lambda_{r}:=\sqrt{(2\lambda_{1})^{2}+\lambda_{2}^{2}},~~\lambda_{a}:=\arctan\left(-\frac{\lambda_{2}}{2\lambda_{1}}\right)\nonumber\\
\rightarrow&\ddot{\Theta}(t)+\lambda_{r}\sin \Theta(t)=0,
\label{eq:equations}
\end{align}
where $\Theta=2\theta'+\lambda_{a}$.
This is nothing but the dynamical equation for a simple pendulum without the approximation of sufficiently small $\Theta$.
The constraints are actually equivalent to
\begin{equation}
\int^{T}_{0}dt \cos \Theta=0,~~\int^{T}_{0}dt \sin \Theta=0.
\label{eq:constraints}
\end{equation}
Owing to the Lipschitz-continuity of the equations, if we can find the solution of Eqs.~(\ref{eq:equations}) and (\ref{eq:constraints}) with taking appropriate multipliers $\lambda_{r}$ and $\lambda_{a}$, it is the unique solution of the original equations and the above an\"{a}tze (a) and (b) are justified.
If the ans\"{a}tze are invalid, the equations return no answer.

The general solution of the above equation is
\begin{equation}
\theta(t)=-\lambda_{a}+\Theta(t),~~\Theta(t)=2 \am(b t+c,k)
\end{equation}
where $\am$ is the Jacobi's amplitude function and $b$, $c$, and $k$ are determined by the boundary conditions and the multipliers (or equivalently, constraints).
Detailed derivations are shown in Appendix \ref{sec:jacobi}.
We then exploit the ansatz (b).
The symmetry of the control field implies 
\begin{equation}
\theta(t)+\theta(T-t)=\theta_{f},
\end{equation}
as Eq.~(\ref{eq:symmetric}), and one can find $\lambda_{a}=-\theta_{f}/2$.
Following the same discussions as those in Sec. \ref{sec:III:pulse},
we obtain Eq.~(\ref{eq:solution2}).

\section{Derivation of the solution}
\label{sec:jacobi}
Let us explain the derivation of the solution of Eq.~(\ref{eq:general_eq}) or (\ref{eq:equations}).
We show again the equation to be solved:
\begin{equation}
\ddot{\Theta}(t)+\sin\left(\Theta(t)\right)=0,
\end{equation}
where $D$ (or $\lambda_{r}$) is dropped by re-scaling the time variable: $s=Dt$ ($\sqrt{\lambda_{r}}t$),
and thus $\dot{\Theta}$ now means $d \Theta/d s$.
Multiplying this by $2 \dot{\Theta}(t)$, we obtain
\begin{align}
&\frac{d}{d s}\left(\dot{\Theta}^{2}(t)-2 \cos\left(\Theta(t)\right)\right)=0\nonumber\\
&\rightarrow \dot{\Theta}^{2}(t)-2 \cos\left(\Theta(t)\right)=E,
\end{align}
where $E$ is the energy of the dynamics (up to a constant factor).
We further deform the equation as
\begin{equation}
\frac{d\left(\Theta/2\right)}{\sqrt{1-k\sin^{2}\left(\Theta/2\right)}}=\pm  \frac{d s}{\sqrt{k}},~~k=\frac{4}{E+2}.
\end{equation}
Integrating both side, and noting that the lhs is the elliptic integral, we obtain 
\begin{align}
\int \frac{d\left(\Theta/2\right)}{\sqrt{1-k\sin^{2}\left(\Theta/2\right)}}=& a t + b,\nonumber\\
\rightarrow\Theta(t)=&2\am\left(a t + b,k\right),
\end{align}
where $a=\pm \sqrt{D/k}$, and $b$ are arbitrary constants.

\section{solution of Eqs. (\ref{eq:time-optimal})}
\label{app:corpse}

In the main text, we only show that the short-CORPSE satisfies Eqs.~(\ref{eq:time-optimal}).
Here we discuss how to derive the solution without such an inference.
We first assume the ansatz (b) in the same way as the case of the pulse-area optimization, which implies that $\lambda=\theta_{f}/2$, $\Theta(0)=-\theta_{f}/2$, and $\Theta(T)=\theta_{f}/2$.

\begin{figure}[h]
		\begin{center}
			\includegraphics[width=170mm]{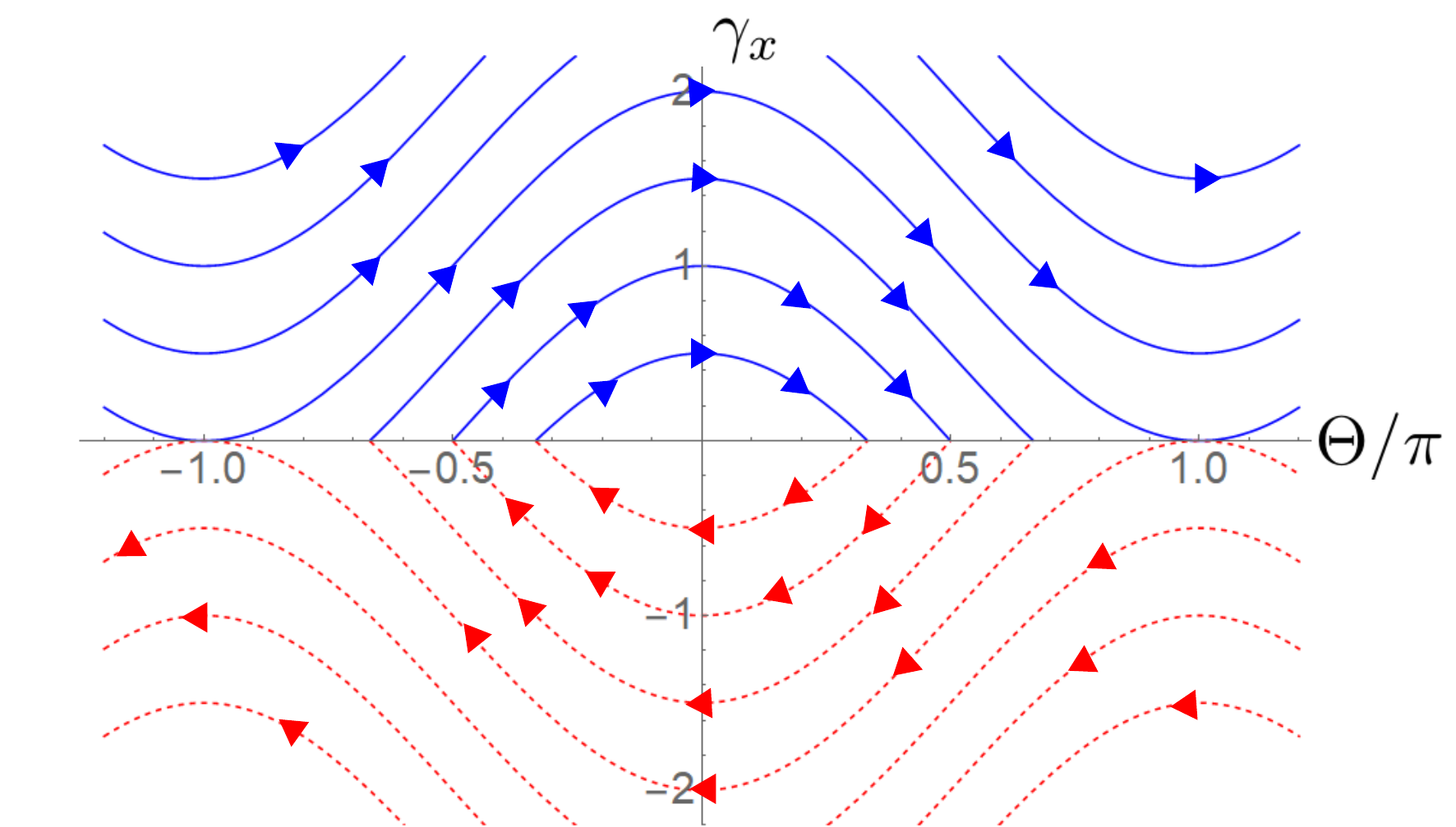}
			\caption{The phase portrait in the $(\Theta,\gamma_{x})$ space.
			The flows in the region $\gamma_{x}>0$ are drawn with blue solid curves while those in the region $\gamma_{x}<0$ are drawn with red dashed curves.
			\label{fig:app_portrait}
			}
		\end{center}
\end{figure}
We then solve $\gamma_{x}(t)$ as a function of $\Theta(t)$ regardless of the constraint term:
\begin{equation}
\gamma_{x}(\Theta)= 
\begin{cases}
D(b+\cos\Theta),~~\gamma_{x}>0,\\
-D\left(b+\cos\Theta\right),~~\gamma_{x}<0.
\end{cases}
\end{equation}
where $b$ and $D$ are constants. $b$ is determined to satisfy the constraint at the end.
On the other hand, $D$ remains arbitrary due to the homogeneity of Eq.~(\ref{eq:timeoptimal_eq_original}) with respect to $\gamma_{I,x,y,z}$ and $\delta_{I,x,y,z}$; hereinafter, we take $D=1$.
When $\gamma_{x}>0$, $\Theta(t)$ increases with a constant speed $\dot{\Theta}=1$ ($\Omega$ in the physical time); when not, $\Theta(t)$ decreases with $\dot{\Theta}=-1$.
We can now draw the phase portrait in the $(\Theta,\gamma_{x})$ plane as in Fig.~\ref{fig:app_portrait}.
The solution is a trajectory from $\left(\Theta(0),\gamma_{x}(0)\right)=\left(-\theta_{f}/2,\pm \left(b+\cos\left(\theta_{f}/2\right)\right)\right)$ to $\left(\Theta(T),\gamma_{x}(T)\right)=\left(\theta_{f}/2,\pm \left(b+\cos\left(\theta_{f}/2\right)\right)\right)$, following to the flow in the portrait.
The sign of $\gamma_{x}(t)$ and $b$ are chosen so that the constraint is satisfied.
Note that $b$ is a constant of motion: when the state $\left(\Theta(t),\gamma_{x}(t)\right)$ transfers at $\gamma_{x}=0$ from the flow in $\gamma_{x}< 0$ to that in $\gamma_{x}> 0$ (or vise versa), the transfer between a mirror pair, which shares the same $b$, is only allowed.

\begin{figure}[h]
		\begin{center}
			\includegraphics[width=170mm]{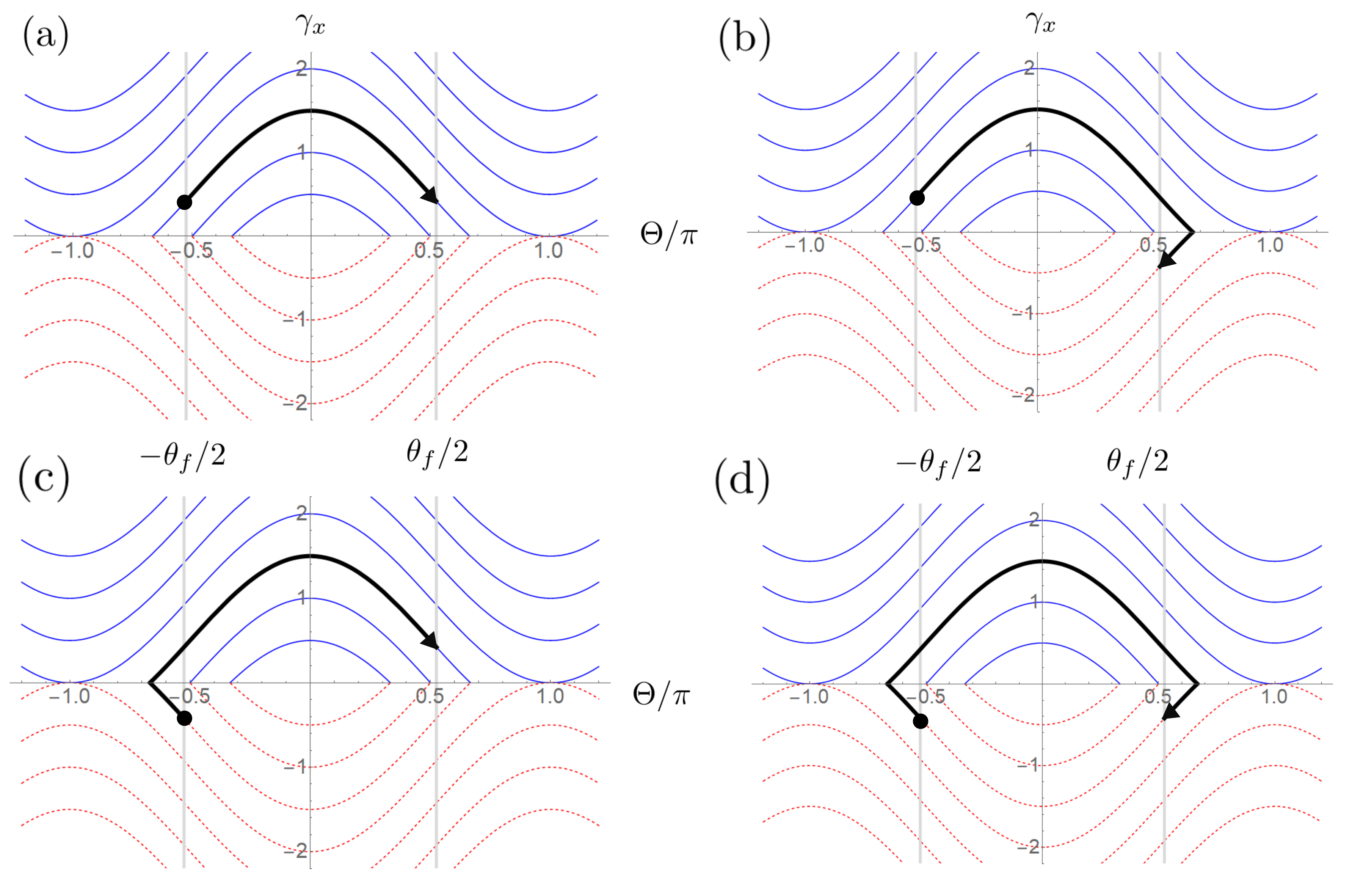}
			\caption{The possible solutions in the phase space.
			The black curve represents the trajectory corresponding to the initial and final states (a), (b), (c), and (d) while the gray lines exhibit $\Theta/\pi=\pm \theta_{f}/2$.
			\label{fig:app_portrait2}
			}
		\end{center}
	\end{figure}

If $|b|>1$, the state cannot move between the two regions.
This implies that $\dot{\Theta}(t)$ does not change its sign, and the control is just a simple rotation, which cannot be detuning-robust.
Thus, $b$ should be less than $1$.
Thanks to the conservation of $b$, the solution is represented by either trajectory in Fig.~\ref{fig:app_portrait2}, which corresponds to the following initial and final states:
\begin{enumerate}[(a).]
\item $\left(\Theta(0),\gamma_{x}(0)\right)=\left(-\theta_{f}/2,k_{b}\right)\rightarrow\left(\Theta(T),\gamma_{x}(T)\right)=\left(\theta_{f}/2,k_{b}\right)$,
\item $\left(\Theta(0),\gamma_{x}(0)\right)=\left(-\theta_{f}/2,k_{b}\right)\rightarrow\left(\Theta(T),\gamma_{x}(T)\right)=\left(\theta_{f}/2,-k_{b}\right)$,
\item $\left(\Theta(0),\gamma_{x}(0)\right)=\left(-\theta_{f}/2,-k_{b}\right)\rightarrow\left(\Theta(T),\gamma_{x}(T)\right)=\left(\theta_{f}/2,k_{b}\right)$, or
\item $\left(\Theta(0),\gamma_{x}(0)\right)=\left(-\theta_{f}/2,-k_{b}\right)\rightarrow\left(\Theta(T),\gamma_{x}(T)\right)=\left(\theta_{f}/2,-k_{b}\right)$,
\end{enumerate}
where $k_{b}=\cos\left(\theta_{f}/2\right)+b$.
Here we have presumed that the solution has no redundant lap, which trivially provides a longer operation time.

The trajectory (a) is nothing but the direct operation, which is not detuning-robust.
The trajectories (b) and (c) have only one change in the sign of $\gamma_{x}$, i.e., the speed $\dot{\Theta}$ changes from $\pm 1$ to $\mp 1$ only one time throughout the dynamics.
As shown in Ref.~\cite{bando2012concatenated}, such a control can neither attain the robustness against the detuning nor amplitude errors.
Thus, the solution has the trajectory (d).
The remaining problem is to solve $b$ whereby the fourth equation (constraint) in Eqs. (\ref{eq:time-optimal}) is satisfied.
The latter constraint including $\sin \Theta$ is trivially satisfied due to anti-symmetry of the trajectory, cf. Eq.~(\ref{eq:symmetry}).
To calculate the former constraint with $\cos \Theta$, we define $(-)\Theta_{\rm SB}$, at which $\gamma_{x}$ changes from positive (negative) to negative (positive), and change the integral variable from $t$ to $\Theta$.
Note that $\Theta$ is a linear function of $t$ and hence it is easy to change variables.
We then obtain,
\begin{align}
\int^{T}_{0}\cos(\Theta) d t=&-\int^{-\Theta_{\rm SB}}_{-\theta_{f}/2} \cos \Theta d \Theta+ \int^{\Theta_{\rm SB}}_{-\Theta_{\rm SB}} \cos \Theta d \Theta-\int^{\theta_{f}/2}_{\Theta_{\rm SB}} \cos \Theta d \Theta =
4\sin \Theta_{\rm SB}-2\sin\left(\theta_{f}/2\right)=0,\nonumber\\
\therefore &~~ \Theta_{\rm SB}=\arcsin\left(\sin\left(\theta_{f}/2\right)/2\right)
\end{align}
As $\gamma_{x}(\Theta_{\rm SB})=0$, one can show that,
\begin{equation}
b=\cos\left(\arcsin\left(\sin(\theta_{f}/2)/2\right)\right)=\cos\kappa,
\end{equation}
which reproduces the short-CORPSE in the main text.
Thus, we have shown that the short-CORPSE is the unique and shortest solution of Eqs. (\ref{eq:time-optimal}).

It should be emphasized again that although the short-CORPSE is the solution of Eqs. (\ref{eq:time-optimal}), it does not guarantee that the short-CORPSE is time-optimal robust control.
The Lipschitz discontinuity makes a gap between the original equations (\ref{eq:timeoptimal_eq_original}) and the reduced ones.

\bibliography{OREoptimal}

\end{document}